\documentclass[prd,aps,floats,twocolumn,showpacs]{revtex4}

\usepackage[dvips]{graphicx}
\usepackage{color}
\usepackage{psfrag}
\definecolor{cyan}{cmyk}{1.,0.,0.,0.5}
\definecolor{vert}{cmyk}{0.5,0.,0.5,0.5}
\definecolor{magenta}{cmyk}{0.,1.,0.,0.5}
\definecolor{verdatre}{cmyk}{0.5,0.,0.5,0.5}
\definecolor{yellow}{cmyk}{0.,0.,1.,0.0}
\definecolor{rouge}{cmyk}{0.,0.4,0.6,0.0}
\definecolor{orange}{cmyk}{0.,0.5,0.5,0.}
\definecolor{violet}{rgb}{0.5,0.,0.5}
\newcommand{\beq}{\begin{equation}}
\newcommand{\eeq}{\end{equation}}
\newcommand{\pbar}{\mbox{$\bar{\rm p}$}}
\begin{document}
\title{The galactic antiproton spectrum at high energies:\\ background expectation vs.
exotic contributions}
\author{Torsten Bringmann}
\email{bringman@sissa.it}
\affiliation{SISSA/ISAS, via Beirut 2-4, 34013 Trieste, Italy}
\author{Pierre Salati} 
\email{salati@lapp.in2p3.fr} 
\affiliation{Universit\'e de Savoie, 73011 Chamb\'ery, France}
\affiliation{Laboratoire de Physique Th\'eorique LAPTH, 74941 Annecy-le-Vieux, France}
\vspace{5ex}
\date{01.05.2007}
\pacs{98.70.Sa, 96.50.sb, 95.35.+d}

\begin{abstract}
A new generation of upcoming space-based experiments will soon start to probe the
spectrum of cosmic ray antiparticles with an unprecedented accuracy and, in particular,
will open up a window to energies much higher than those accessible so far. It is thus
timely to carefully investigate the expected antiparticle fluxes at high energies. Here,
we perform such an analysis for the case of antiprotons. We consider both standard
sources as the collision of other cosmic rays with interstellar matter, as well as exotic
contributions from  dark matter annihilations in the galactic halo. Up to energies well
above 100 GeV, we find that the background flux in antiprotons is almost uniquely
determined by the existing low-energy data on various cosmic ray species; for even higher
energies, however, the uncertainties in the parameters of the underlying propagation
model eventually become significant. We also show that if the dark matter is composed of
particles with masses at the TeV scale, which is naturally expected in extra-dimensional
models as well as in certain parameter regions of supersymmetric models, the annihilation
flux can become comparable to -- or even dominate -- the antiproton background at the
high energies considered here.
\end{abstract}
\maketitle

%%%%%%%%%%%%%%%%%%%%%%%%%%%%%%%%%%%%%%%%%%%%%%%%%%%%%%%%%%%%%%%%%%%%%%%%%%%%%%
\newcommand{\be}{\begin{equation}}
\newcommand{\ee}{\end{equation}}
\newcommand{\bea}{\begin{eqnarray}}
\newcommand{\eea}{\end{eqnarray}}
\newcommand{\B}{{B^{(1)}}}

%%%%%%%%%%%%%%%%%%%%%%%%%%%%%%%%%%%%%%%%%%%%%%%%%%%%%%%%%%%%%%%%%%%%%%%%%%%%%%
\section{Introduction}

Since the pioneering works of Victor Hess almost 100 years ago we know that the earth is
constantly exposed to a bombardment of cosmic rays with presumably galactic origin.
Extending up to extremely high energies, they mainly consist of protons (90\%) and helium
(9\%), but also contain sizeable contributions of other nuclei and antiparticles. Most of
these particles have probably been accelerated in the blast waves of supernova remnants,
while others -- like antiprotons -- are rather produced by the spallation of the
interstellar medium by primary cosmic rays. 

Measurements of the cosmic-ray antiproton spectrum started in the late 1970s and now
extend up to energies around 30 GeV \cite{bess1,bess2,caprice,ams98}.
The observations can be extremely well described by adopting a production mechanism
as indicated above and then propagating the antiprotons through the
galactic halo by means of a simple diffusion model \cite{don01}. Remarkably, the
parameters of this model are essentially fixed by the spectra of other cosmic ray
species, in particular the boron over carbon (B/C) ratio \cite{mau01}. With the
satellite-borne PAMELA experiment \cite{PAM} being in orbit since June 2006 and already taking data, and the planned installation of AMS-02 \cite{AMS} on the international
space station by 2007, the near future will witness very
accurate and detailed measurements of the antiproton spectrum, in the case of AMS-02 up
to energies of around 1 TeV. It is therefore of great interest to extend previous
analyses of the background spectrum to higher energies since this will provide an
important test of the underlying diffusion model and thus of our understanding of the
propagation of cosmic rays through the galaxy. 
Having this in mind, the first part of the present work is devoted to a careful
discussion of the uncertainties in the expected background spectrum at high energies and,
as we shall see, how the already available data help to constrain it considerably.

Not only is a thorough understanding of the cosmic ray antiproton background an
interesting issue in itself, it is also a mandatory prerequisite for any attempt to spot
possible exotic contributions. The main reason to expect these is the by now
well-established existence of dark matter (DM) which, from the most recent measurements
of the cosmic microwave background, contributes about 23 \% to the total energy content
of the universe \cite{WMAP}. While the nature of dark matter is still an open question,
weakly interacting massive particles (WIMPs) that arise in many extensions to the
standard model of particle physics (SM) are very plausible candidates; thermally produced in
the early universe, they automatically acquire the right relic density today
\cite{WIMPDM}. As has first been noticed in \cite{sil84}, the self-annihilation of these
particles in the galactic halo could then leave an imprint in the antiproton spectrum.
Traditionally, one has mainly focused on a possible contribution at rather low energies
since first measurements seemed to indicate an excess of antiprotons below the peak at
around 1 GeV \cite{buf81}. With the improved statistics of follow-up experiments and a
better understanding of the production mechanisms, however, the evidence for such an
excess disappeared \cite{ber99}. In this article, we will therefore rather focus on
possible exotic signatures at the high energies that soon will be accessible for the
first time. To this end, we will compare the situation for three benchmark models of
realistic dark matter candidates with masses in the TeV range; the lightest Kaluza-Klein
particle (LKP) in models with universal extra dimension \cite{ued,LKP} and the lightest
supersymmetric particle (LSP) in the limit where it is a neutralino with a very high
Higgsino \cite{Higgsino} or Wino \cite{his04} fraction, respectively. For earlier work
that explicitly studies the effect of dark matter annihilations on the high-energy
antiproton spectrum, see e.g. \cite{ull99,lio05,bri05,his05}.

This article is organized as follows. In Section \ref{sec_prop}, we briefly review the
two-zone diffusion model that is used to propagate charged particles through the galactic
halo; as mentioned before, this model provides an excellent fit to the low-energy data.
We then continue in Section \ref{sec_background} with a detailed discussion, both
numerically and analytically, of the expected antiproton background for energies up to 10
TeV. Section \ref{sec_signal} introduces the benchmark dark matter models that then will
be used to discuss possible signatures for exotic physics in the antiproton flux and the
prospects for their detection. In Section \ref{sec_conc}, finally, we give a short
summary of our results and present our conclusions.

%%%%%%%%%%%%%%%%%%%%%%%%%%%%%%%%%%%%%%%%%%%%%%%%%%%%%%%%%%%%%%%%%%%%%%%%%%%%%%
%%%%%%%%%%%%%%%%%%%%%%%%%%%%%%%%%%%%%%%%%%%%%%%%%%%%%%%%%%%%%%%%%%%%%%%%%%%%%%
\section{Propagation of charged particles through the galaxy}
\label{sec_prop}

%Diffusion equation for charged particles in general.
%
{
Whatever the mechanism responsible for their production, charged cosmic
rays subsequently propagate through the galactic magnetic field and bounce off its
irregularities -- the Alfv\'en waves. The resulting particle transport  is well
described by space diffusion with a coefficient
\be
 K(E) = K_{0} \, \beta \, {\mathcal R}^{\delta}\,,
\ee
 which increases as a
power law with the rigidity ${\mathcal R}$ of the particle.
In addition, since these scattering centers move  with a velocity
$V_{a} \sim$ 20 to 100 km s$^{-1}$, a second order Fermi mechanism is 
responsible for some diffusive re-acceleration. Its coefficient $K_{EE}$
depends on the particle velocity $\beta$ and total energy $E$ and is
related to the space diffusion coefficient $K(E)$ through
\beq
K_{EE} \; = \;
{\displaystyle \frac{2}{9}} \, V_{a}^{2} \,
{\displaystyle \frac{E^{2} \beta^{4}}{K(E)}} \;\; .
\eeq
In the case of nuclei and antiprotons, one also has to take into account that a certain energy loss rate
$b^{\rm loss}(E)$ is induced by adiabatic and Coulomb
energy losses, as well as by ionization. Finally, galactic convection wipes cosmic rays away
from the disk with a velocity $V_{C} \sim$ 5 to 15 km s$^{-1}$.

After these preliminaries, and equipped with all the necessary notation, we may now write the master equation for the distribution function $\psi = dn / dE$ in space and in energy as:
\be
\partial_{z} \left( V_{C} \, \psi \right)  -  K  \Delta \psi %\hspace{30ex}\nonumber\\
+\partial_{E} \left\{ b^{\rm loss}(E)  \psi  - 
K_{EE}(E)  \partial_{E} \psi \right\}  =  q\,.
\label{master_equation}
\ee
This equation applies to any species -- protons, antiprotons or positrons
-- as long as the rates for production $q$ and energy loss $b^{\rm loss}(E)$
are properly accounted for.
This general framework -- summarized in Eq.~(\ref{master_equation}) -- may be
implemented within the semi-analytical two-zone model which is extensively
discussed in~\cite{mau01,don01} and whose salient features are briefly recalled
here.}

{
In this approach, a steady state is assumed and the diffusive halo is pictured as a thick wheel which matches
the circular structure of the Milky Way. The galactic disk of stars and gas --
where primary cosmic rays are accelerated -- lies in the middle. It extends
radially 20 kpc from the center and has a half-thickness $h$ of 100 pc. Confinement
layers where cosmic rays are trapped by diffusion lie above and beneath that disk.
The intergalactic medium starts at the vertical boundaries $z = \pm L$ as well as
beyond a radius of $r = R \equiv 20$ kpc. Notice that the half-thickness $L$
of the diffusive halo is not known and reasonable values range from 1 to 15 kpc.
The diffusion coefficient
$K$ is the same everywhere whilst the convective velocity is exclusively vertical
with component $V_{C}(z) = V_{C} \, {\rm sign}(z)$. The galactic wind -- produced
by the bulk of the disk stars like the Sun -- drifts away from them along the
vertical directions, hence the particular form assumed here for $V_{C}$.
The spallation of the interstellar gas by cosmic rays takes place in the disk.
It leads to the production of secondary species like boron or antiprotons.
The latter is a background against the neutralino annihilation signal and will
be discussed in section~\ref{sec_background}.}

%General solution in terms of Bessel expansion.
%

{
The diffusive halo is axisymmetric and the cosmic ray density vanishes at the
radius $R = 20$ kpc. This condition is naturally implemented by the following series expansion for $\psi$:
\beq
\psi \left( r , z , E \right) \; = \;
{\displaystyle \sum_{i=1}^{+ \infty}} \; P_{i} \left( z , E \right) \,
J_{0} \left( \alpha_{i} \, r / R \right) \, .
\label{bessel_psi}
\eeq
The Bessel function of zeroth order $J_{0}$ vanishes at the points $\alpha_{i}$.
The radial dependence of $\psi$ is now taken into account by the set of its Bessel
transforms $P_{i}(z,E)$. The source term $q$ may also be Bessel expanded into
the corresponding functions $Q_{i}(z,E)$ and the master
equation~(\ref{master_equation}) thus becomes
\begin{eqnarray}
\partial_{z} \left( V_{C} \, P_{i} \right)  - 
K  \partial_{z}^{2} P_{i}  + 
K  \left\{ {\displaystyle \frac{\alpha_{i}^{2}}{R^{2}}} \right\}  P_{i}
   \label{master_2}\hspace{18ex}\\
+{2  h \, \delta(z)} \,
\partial_{E} \left\{ b^{\rm loss}(E) \, P_{i}  - 
K_{EE}(E) \, \partial_{E} P_{i} \right\}=Q_{i} \left( z , E \right)
\, . \nonumber\hspace{-3ex}
\end{eqnarray}
Here, energy loss and diffusive reacceleration are confined inside the galactic disk --
which is considered infinitely thin, hence the presence of an effective term
$2 h \, \delta(z)$.

The form of the source term
$Q_{i}(z,E)$ that appears in  Eq.~(\ref{master_2}) depends on
the type of the particle species that is considered.
In the case of antiprotons, the following mechanisms can in principle contribute:

\begin{itemize}
\item
Antiprotons may collide elastically on interstellar H and He. Because they
are preferentially scattered forward, however, such interactions are innocuous
and will be disregarded.

\item
Antiprotons may also annihilate on interstellar H and He. This leads to
a negative source term $- \, \Gamma_{\pbar}^{\rm ann} \, \psi$, where
the annihilation rate $\Gamma_{\pbar}^{\rm ann}$ is defined as
\beq
\Gamma_{\pbar}^{\rm ann} \; = \;
\sigma_{\pbar \, {\rm H}}^{\rm ann}  \, v_{\pbar} \, n_{\rm H} \, + \,
\sigma_{\pbar \, {\rm He}}^{\rm ann} \, v_{\pbar} \, n_{\rm He} \, .
\eeq
The annihilation cross section $\sigma_{\pbar \, {\rm H}}^{\rm ann}$ is
borrowed from \cite{Tan_Ng_82,Tan_Ng_83} and we have multiplied it by a
factor of $4^{2/3} \sim 2.5$, taking into account the higher geometric cross section, to get $\sigma_{\pbar \, {\rm He}}^{\rm ann}$.
The average hydrogen and helium densities in the galactic disk have been
set equal to $n_{\rm H} = 0.9$ cm$^{-3}$ and
$n_{\rm He} = 0.1$ cm$^{-3}$, respectively.

\item
The annihilation of DM candidate particles throughout the Milky Way halo
generates primary antiprotons. The corresponding source term
$q_{\pbar}^{\rm prim} \left( r , z , E \right)$ will be discussed in
section~\ref{sec_signal}. Notice that annihilations take place all over
the diffusive halo.

\item
The latter is not the case neither for secondary antiprotons -- which are
produced as high energy primary nuclei impinge on the atoms of the
interstellar medium inside the galactic disk -- nor for tertiary
antiprotons which result from the inelastic and non-annihilating
interactions which these particles may undergo with the same atoms.
Antiprotons may actually collide on a proton at rest and transfer
enough energy to excite it as a $\Delta$ resonance. This mechanism
redistributes antiprotons towards lower energies and flattens their
spectrum \cite{ber99}.

\end{itemize}

The rate for the production of secondary antiprotons, $q_{\pbar}^{\rm sec}(r , E_{\pbar})$, is discussed in more detail in Section \ref{sec_background}; for tertiary antiprotons, it is given by:
\begin{eqnarray}
&&\hspace{-3ex}q_{\pbar}^{\rm ter}(r , E_{\pbar})=\nonumber\\ 
&&{\displaystyle \int_{E_{\pbar}}^{+ \infty}} \,
{\displaystyle
\frac{d \sigma_{\rm \bar{p} \, H \to \bar{p} \, X}}{dE_{\pbar}}}
\left( E'_{\pbar} \to E_{\pbar} \right) \; n_{\rm H} \;
v'_{\pbar} \; \psi_{\pbar} (r,E'_{\pbar}) \;
dE'_{\pbar} \nonumber \\
&& - 
\sigma_{\rm \bar{p} \, H \to \bar{p} \, X}
\left( E_{\pbar} \right) \, n_{\rm H} \;
v_{\pbar} \; \psi_{\pbar} (r,E_{\pbar}) \;\; ,
\label{tertiary}
\end{eqnarray}
where the inelastic and non-annihilating differential cross-section
in this expression can be approximated by
\begin{equation}
{\displaystyle
\frac{d \sigma_{\rm \bar{p} \, H \to \bar{p} \, X}}{dE_{\bar{\rm p}}}}
\, = \, {\displaystyle
\frac{\sigma_{\rm \bar{p} \, H \to \bar{p} \, X}}{T'_{\bar{\rm p}}}} \,.
\end{equation}
Here, $T'_{\bar{\rm p}}$ is the initial antiproton kinetic energy. In order to take into account elastic scattering on helium, one simply has to replace the hydrogen density by
$n_{\rm H} \, + \, 4^{2/3} \, n_{\rm He}$.

\begin{widetext}
With all these source terms specified, the full expression for the master 
equation describing the (Bessel transformed) antiproton distribution function $\bar{P}_{i}(z,E)$ becomes:
\begin{eqnarray}
\partial_{z} \left( V_{C} \, \bar{P}_{i} \right) \; - \;
K \, \partial_{z}^{2} \bar{P}_{i} \; + \;
K \, \left\{
{\displaystyle \frac{\alpha_{i}^{2}}{R^{2}}} \right\} \, \bar{P}_{i}
& + &
2 \, h \, \delta(z) \,
\partial_{E} \left\{ b^{\rm loss}(E) \, \bar{P}_{i} \; - \;
K_{EE}(E) \, \partial_{E} \bar{P}_{i} \right\} \; = \; \nonumber \\
& - & \, 2 \, h \, \delta(z) \,
\Gamma_{\pbar}^{\rm ann} \, \bar{P}_{i} \; + \;
Q_{\pbar , i}^{\rm prim} \left( z , E \right) \; + \;
2 \, h \, \delta(z) \,
\left\{ Q_{\pbar , i}^{\rm sec} + Q_{\pbar , i}^{\rm ter} \right\} \;\; .
\label{master_susy}
\end{eqnarray}
Integrating this relation along the vertical axis $z$
-- in particular through the infinitely thin disk -- finally leads to a diffusion
equation in energy which the Bessel transforms
$\bar{P}_{i} \left( 0 , E \right)$ fulfill:

\begin{eqnarray}
\bar{\cal A}_{i} \, \bar{P}_{i} \left( 0 , E \right) \; + \;
2 \, h \; \partial_{E} \left\{ b^{\rm loss}(E) \, \bar{P}_{i} \left( 0 , E \right)
\right. & \! - & \! \left.
K_{EE}(E) \, \partial_{E} \, \bar{P}_{i} \left( 0 , E \right) \right\}
\; = \; \nonumber \\
&  & 2 \, h \,
\left\{ Q_{\pbar , i}^{\rm sec} + Q_{\pbar , i}^{\rm ter} \right\}
\; + \;
2 \,
{\displaystyle \int_{0}^{L}} \, dz \;
Q_{\pbar , i}^{\rm prim} \left( z , E \right) \;
e^{- \, {\displaystyle \frac{V_{C} z}{2 K}}} \;
{\mathcal F}_{i}(z) \;\; .
\label{master_final}
\end{eqnarray}
\end{widetext}
The coefficients $\bar{\cal A}_{i}$ that appear in the above expression are given by
\beq
\bar{\cal A}_{i}(E) = V_{C} \, + \,
2 \, h \, \Gamma_{\pbar}^{\rm ann}(E) \, + \,
K(E) \, S_{i} \coth \left( {\displaystyle \frac{S_{i} \, L}{2}} \right) ,
\eeq
where $S_{i}^{2} \, = \, ({V_{C}}/{K})^{2} \, + \, ({2 \, \alpha_{i}}/{R})^{2}$,
and the vertical functions ${\mathcal F}_{i}(z)$ are defined as
\beq
{\mathcal F}_{i}(z) \; = \;
{\sinh \left\{ {\displaystyle \frac{S_{i}}{2}} \, \left( L - z \right) \right\}}
\, / \,
{\sinh \left\{ {\displaystyle \frac{S_{i}}{2}} \, L \right\}} \;\; .
\label{F_de_i}
\eeq
In this work, we solve Eq.~(\ref{master_final}) by following the method
explained in the appendix B of \cite{don01}.
}

%Green function for antiprotons.
%

{
A completely different approach to describe the antiproton propagation through the diffusive halo relies on the existence of a Green function
$G_{\pbar}$. Such a function translates the
probability for an antiproton produced at point $S(x_{S},y_{S},z_{S})$
to travel to the observer located at point $M(x,y,z)$. The antiproton
energy spectrum is given by the convolution of the Green function $G_{\pbar}$
with the production rate $q_{\pbar}$
\beq
\psi(M,E) \; = \; {\displaystyle \int} \, d^{3}{\mathbf x}_{S} \;\;
G_{\pbar}(M \leftarrow S , E) \, q_{\pbar}(S,E) \;\; .
\label{convolution_1}
\eeq
Energy loss, diffusive reacceleration and tertiary production are
inefficient above a few GeV. For our purposes, we may therefore neglect these processes
when deriving the antiproton propagator $G_{\pbar}$.
The Milky Way is now pictured as an infinite slab of
half-thickness $L$ with a gaseous disk in the middle at $z = 0$.
The antiproton propagation is invariant under a translation along
the horizontal axis $x$ or $y$. The master equation~(\ref{master_equation})
needs still to be solved along the vertical direction $z$ with the
condition that $G_{\pbar}$ vanishes at the boundaries $z = \pm L$.
The construction of the Green function for antiprotons is inspired
from the positron case -- see in particular section 3 of \cite{lav06}
-- with the difference that the antiproton energy does not change
and that time is integrated out. Explicit expressions for $G_{\pbar}$
may be found in the appendix C.3.a of \cite{G_pbar_0609522}.
%as well as in \cite{brun06}.
Because we are interested in the flux at the earth
-- i.e.~within the disk, at $z = 0$ -- the integral~(\ref{convolution_1})
may be recast into
\bea
\psi(x_{\odot}=r_{\odot},y_{\odot}=0,z_{\odot}=0,E) \; = \hspace{17ex}\nonumber\\
4 \pi \, {\displaystyle \int_{0}^{L}} \, dz_{S} \, {\displaystyle \int_{0}^{R}} \,
r_{S} \, dr_{S} \;\;
G_{\pbar}(\odot \leftarrow S , E) \, q_{\pbar}(S,E) \, ,
\label{convolution_2}
\eea
where the antiproton propagator simplifies in that case into the series
\bea
G_{\pbar}(\odot \leftarrow S , E) \; = \;\hspace{32ex}\nonumber\\
{\displaystyle \frac{e^{\displaystyle - z_{S} / r_{w}}}{2 \pi K(E)}} \;\;
{\displaystyle \sum_{n=1}^{+ \infty}} \; {\displaystyle \frac{1}{C_{n}}} \;
\phi_{n}(0) \, \phi_{n}(z_{S}) \;
K_{0} \left( {\displaystyle \frac{r}{L}} \, \sqrt{\epsilon_{n}} \right) \, .
\label{G_pbar}
\eea
Here, the vertical functions
$\phi_{n}$  are given by
\beq
\phi_{n}(z) \; = \; \sin
\left\{ \xi_{n} \left( 1 - {\displaystyle \frac{z}{L}} \right) \right\}
\, ,
\eeq
where the coefficients $\xi_{n}$ are solutions to the equation
\beq
\xi_{n} \; = \; n \, \pi \; - \; \tan^{-1} \left( p \, \xi_{n} \right) \, .
\eeq
At a given antiproton energy $E$, two specific scales may be singled out;
the scattering length $r_{s}^{-1} \equiv h\, \Gamma_{\pbar}^{\rm ann}(E) / K(E)$ and
a convective scale $r_{w}^{-1} \equiv V_{C} / 2 K(E)$. 
The parameter
\beq
{\displaystyle \frac{1}{p}} \; \equiv \;
{\displaystyle \frac{L}{r_{s}}} \, + \,
{\displaystyle \frac{L}{r_{w}}} 
\eeq
then serves to compare these two.
At high energies, above $\sim 100$ GeV, $r_{s}$ and $r_{w}$ are much larger than $L$ due to the greatly enhanced diffusion coefficient $K(E)$.
As a consequence, we have $p\gg 1$ and thus
$\xi_{n} \, \approx \, (n - 1/2) \, \pi$.
The scale $C_{n}$, defined by
\beq
{\displaystyle \frac{C_{n}}{L}} \; = \; 1 \, + \,
{\displaystyle \frac{1}{p}}
\left( {\displaystyle \frac{\sin \xi_{n}}{\xi_{n}}} \right)^{2} \, ,
\eeq
tends to $L$ in this regime where $p$ is large, i.e. when diffusion takes over disk annihilations
and galactic convection.
In the argument of the modified Bessel functions of the second kind $K_{0}$ in Eq.~(\ref{G_pbar}),
the ratio $r/L$ is multiplied by a factor $\sqrt{\epsilon_{n}}$ where
\beq
\epsilon_{n} \; = \; \xi_{n}^{2} \, + \,
\left( {\displaystyle \frac{L}{r_{w}}} \right)^{2} \;\; ,
\eeq
which reduces to $\xi_{n}^{2}$ in the pure diffusive regime at high
energy.
}

{
As demonstrated later in more detail, the Bessel expansion and the 
propagator approach give similar results. The small discrepancies
between the two methods will be discussed in section~\ref{sec_signal}
but we may already now safely use Eq.~(\ref{G_pbar}) to get a deeper
insight into antiproton propagation. In particular, we see that at high energies, the
diffusion coefficient $K(E)$ in front of the expansion~(\ref{G_pbar})
is the only term that depends on the energy $E$. In the following Section, we will return to this observation when we discuss the expected antiproton background at high energies. 
}

%%%%%%%%%%%%%%%%%%%%%%%%%%%%%%%%%%%%%%%%%%%%%%%%%%%%%%%%%%%%%%%%%%%%%%%%%%%%%%
%%%%%%%%%%%%%%%%%%%%%%%%%%%%%%%%%%%%%%%%%%%%%%%%%%%%%%%%%%%%%%%%%%%%%%%%%%%%%%
\section{The antiproton background}
\label{sec_background}

%production mechanism for secondary $\bar p$. We specify here the source
%function and discuss the cross section $d\sigma_{CR+ISM \rightarrow \pbar + X}$.
%We start with p + H.
%
{
Secondary antiprotons are produced by the spallation of the interstellar
gas of the Milky Way disk -- mostly hydrogen and helium -- by impinging
cosmic ray primaries. In the case of the interactions between cosmic ray
protons and hydrogen atoms, the source term takes the following form:
\bea
q_{\pbar}^{\rm sec}(r , E_{\pbar}) \; = \;\hspace{35ex}
\label{source_sec}\\
{\displaystyle \int_{E^{0}_{\rm p}}^{+ \infty}} \,
{\displaystyle \frac{d \sigma_{\rm p \, H \to \bar{p}}}{dE_{\pbar}}}
\left\{ E_{\rm p} \to E_{\pbar} \right\} \; n_{\rm H} \;
v_{\rm p} \; \psi_{\rm p} (r,E_{\rm p}) \;
d E_{\rm p} \;\; . \nonumber
\eea
{The proton and helium cosmic ray fluxes at the Earth
have been measured accurately and have been borrowed from \cite{don01}.
For each cosmic ray model, they have been consistently propagated backward in the
whole diffusive halo in order to yield $\psi_{\rm p} (r,E_{\rm p})$ and
$\psi_{\rm He} (r,E_{\rm nuc}=E_{\rm p})$.}

For kinematic reasons, the production rate peaks around a few GeV; this is because the proton
energy must be larger than a threshold of $E^{0}_{\rm p} = 7 \, m_{\rm p}$.
In the galactic frame, the differential production cross section that enters
in the previous relation is given by the integral
\beq
{\displaystyle \frac{d \sigma}{dE_{\pbar}}} \; = \;
2 \, \pi \, k_{\pbar} \,
{\displaystyle \int_{0}^{\theta_{\rm max}}}
\left( E_{\pbar} \,
{\displaystyle \frac{d^{3} \sigma}{d^{3} {k}_{\pbar}}} \right)_{\rm LI}
d \left( - \cos \theta \right) \;\; ,
\label{dif_cross_section}
\eeq
where $\theta$ denotes the angle between the momenta of the incoming proton
and the produced antiproton. In the center of mass frame, which drifts
with a velocity
$\beta_{\rm CM} =
\left\{(E_{\rm p} - m_{\rm p})/(E_{\rm p} + m_{\rm p})\right\}^{1/2}$
with respect to the galactic frame, the antiproton energy cannot exceed
a value of
\beq
E_{\pbar , {\rm max}}^{*} \; = \;
{\displaystyle \frac{s \, - \, 9 \, m_{\rm p}^{2} \, + \, m_{\rm p}^{2}}{2 \sqrt{s}}}
\;\; ,
\eeq
where $\sqrt{s} = \left\{2 m_{\rm p} (E_{\rm p} + m_{\rm p})\right\}^{1/2}$
is the total energy of the reaction. In Eq.~(\ref{dif_cross_section}), the
energies $E_{\rm p}$ and $E_{\pbar}$ have been fixed and the angular integral
runs from $\theta = 0$ up to a maximal value of $\theta_{\rm max}$ for which
\beq
\cos \theta_{\rm max} \; = \;
{\displaystyle \frac{1}{\beta_{\rm CM} \, k_{\pbar}}} \,
\left( E_{\pbar} \, - \,
{\displaystyle \frac{E_{\pbar , {\rm max}}^{*}}{\Gamma_{\rm CM}}} \right) \;\; .
\eeq
The Lorentz invariant differential cross section
$E_{\pbar} \, ({d^{3} \sigma}/{d^{3} {k}_{\pbar}})$ depends
on the antiproton rapidity $y = \tanh^{-1}(k_{\pbar \, \parallel}/E_{\pbar})$
and transverse mass $m_{\rm T}^{2} =  m_{\rm p}^{2} + k_{\pbar \, \perp}^{2}$;
it has been parameterized according to \cite{Tan_Ng_82,Tan_Ng_83}.
In order to avoid numerical problems in the computation of (\ref{dif_cross_section}), it is convenient to change the integration variable from $\theta_{\rm max}$ to $m_{\rm T}$ for small angles $\theta_{\rm max}$.
}

%We discuss nuclei now.
%

{
So far, we have discussed the spallation of hydrogen by high-energy protons.
Let us now turn to helium, which  is also present in both the cosmic radiation as well as in the
interstellar material of the galactic disk.
As regards the spallation of interstellar helium by cosmic ray protons,
the derivation of the differential cross section
${d \sigma}/{dE_{\pbar}}({\rm p \, He \to \bar{p}})$ follows the same lines
as explained above. Because the incoming proton interacts with a single nucleon
of the target nucleus, the center of mass frame of the collision is the same
as before. The Lorentz invariant differential cross section -- which is the
only new ingredient -- has been parameterized in~\cite{duperray03}
 as a function of $\sqrt{s}$, $k_{\pbar , \perp}$ and the ratio
$x_{R} = {E_{\pbar}^{*}}/{E_{\pbar , {\rm max}}^{*}}$. The total inelastic
cross section for $p + A$ collisions is taken from~\cite{letaw83};
a boost is then necessary to cope with the collisions of high-energy helium nuclei
on interstellar hydrogen. An antiproton with rapidity $y$ in the reference
system of the hydrogen target has a rapidity
$- y' \, = \, \cosh^{-1}(E_{\rm nuc}/m_{\rm p}) \, - \, y$ in
the reference frame of the impinging helium nucleus. The integral~(\ref{source_sec})
runs now over the energy per nucleon $E_{\rm nuc}$ of the incident cosmic ray He.
Finally, because helium amounts to a fraction of $\sim$ 0.1 with respect to
hydrogen, we expect He + He interactions to contribute only a few percent to
the secondary antiproton flux. Inspired by~\cite{engel92}, we have
multiplied the differential cross section
${d \sigma}/{dE_{\pbar}}({\rm p \, He \to \bar{p}})$ by the effective
number $\eta$ of nucleons inside the impinging He nucleus that participate
in the reaction. It is given by the ratio of the average number
$<n_{\rm int}>$ of nucleon-nucleon interactions over the average number
$<n_{\rm T}>$ of target nucleons involved. Both $<n_{\rm int}>$ and
$<n_{\rm T}>$ depend on the center of mass energy. At $\sqrt{s} = 10$ GeV
for instance, we have $<n_{\rm int}> = 3.68$ and $<n_{\rm T}> = 2.64$, thus leading to
$\eta = 1.39$. This value increases up to $\eta = {5.12}/{3.25} = 1.58$ at
$\sqrt{s} = 1$ TeV, not too far from the simple scaling factor of
$A^{1/3} = 1.59$ which we have used in our code.
}

\begin{figure}[t!]
\includegraphics[width=\columnwidth]{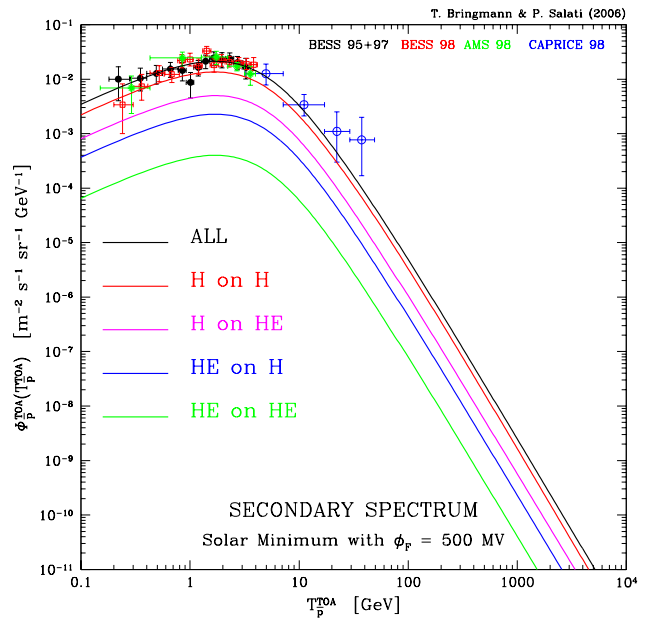}
\caption{The various contributions to secondary antiprotons from the spallation of the
interstellar medium by cosmic rays. Here, we took the `medium' configuration of
propagation parameters from Table~\ref{tab_prop} and $T_{\bar p}^\mathrm{TOA}$ denotes
the antiproton kinetic energy as measured at the top of the atmosphere. For reference, we
also show the existing low-energy data on the antiproton flux at the top of the
atmosphere \cite{bess1,bess2,caprice,ams98}.}
\label{fig_contributions}
\end{figure}

\begin{figure*}[t!]
\begin{minipage}[t]{0.49\textwidth}
\centering
\includegraphics[width=\textwidth]{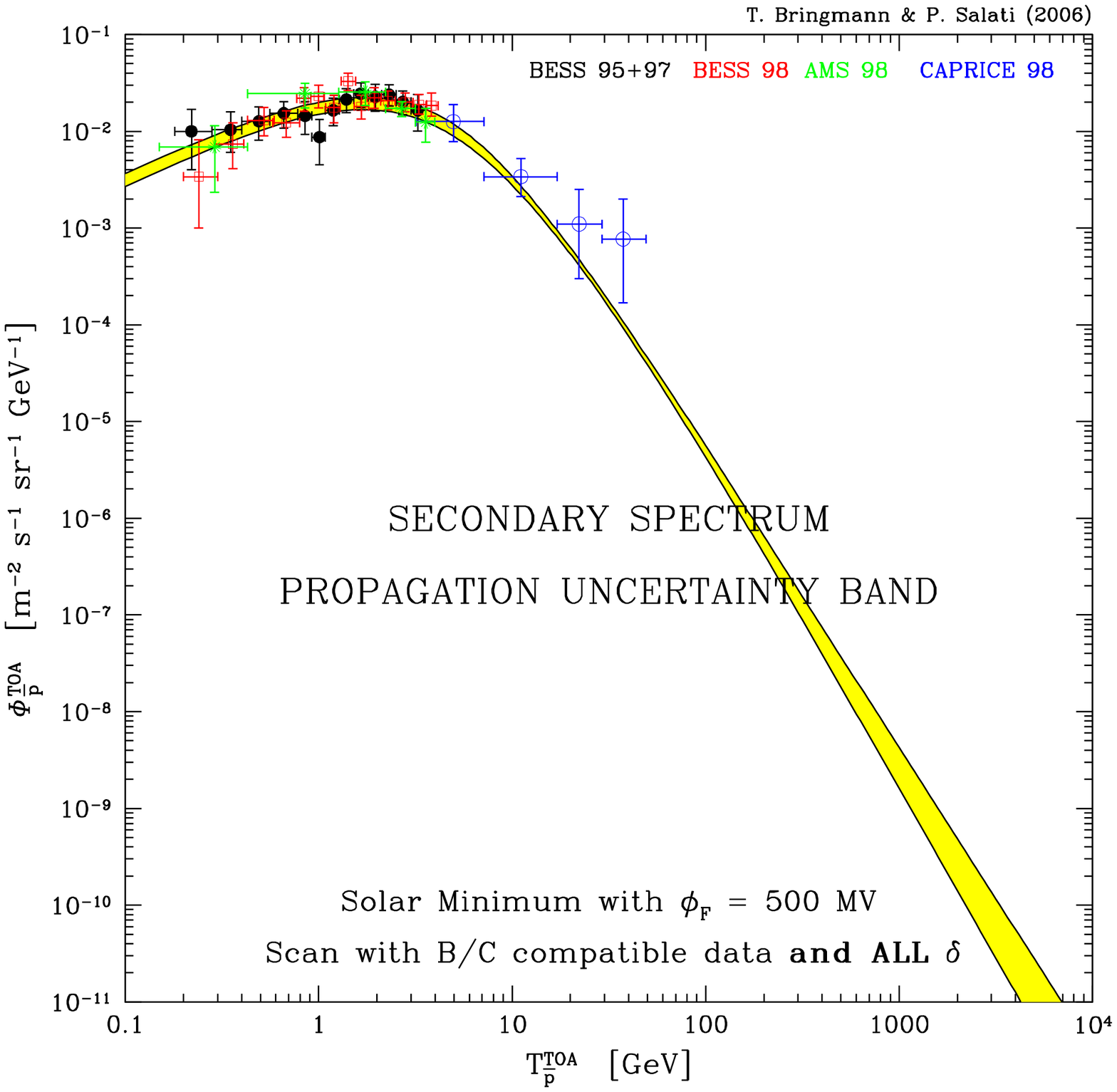}\\
\end{minipage}
\begin{minipage}[t]{0.02\textwidth}
\end{minipage}
\begin{minipage}[t]{0.49\textwidth}
\centering  
\includegraphics[width=\textwidth]{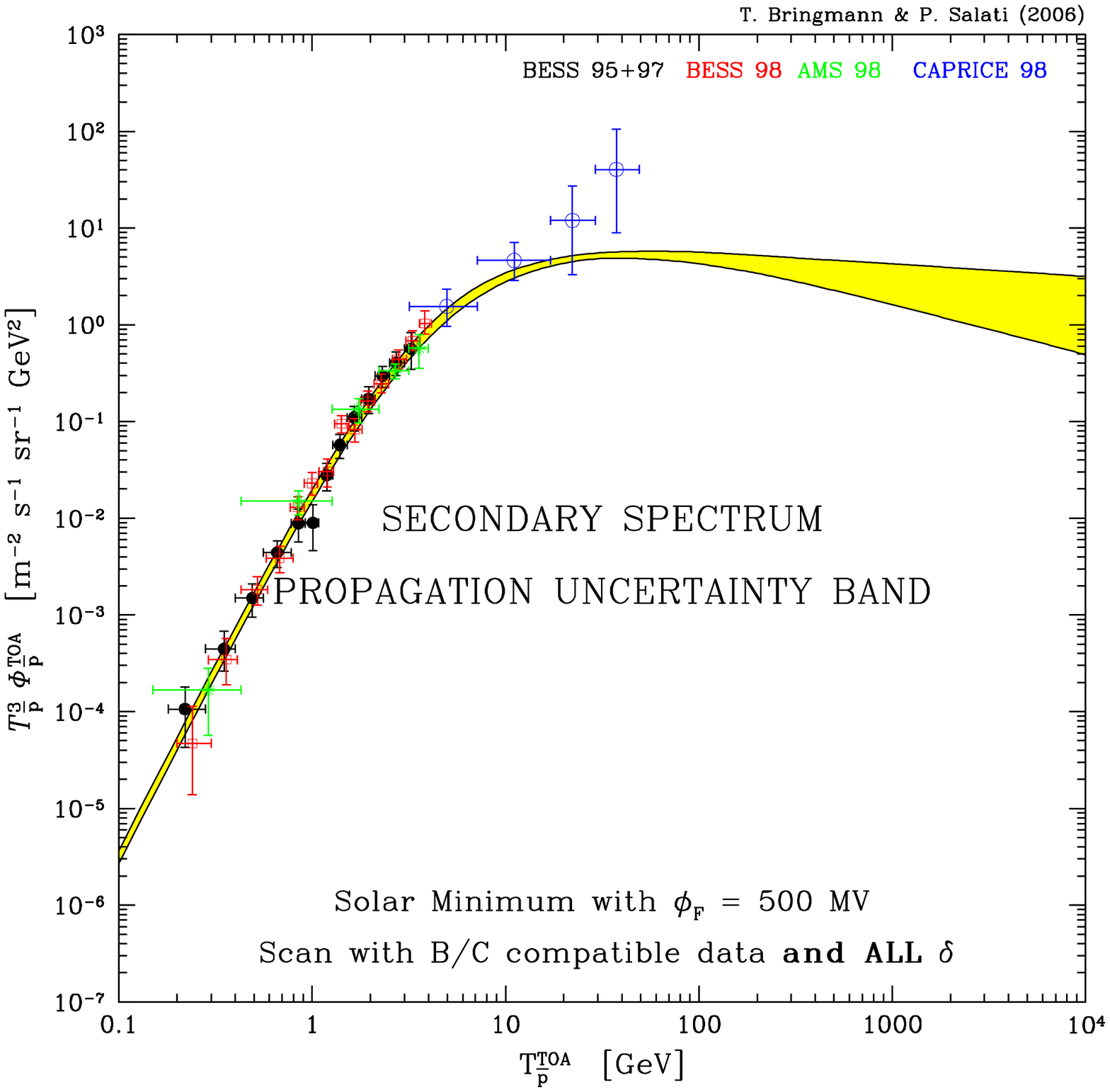}\\
\end{minipage}
\caption{Theoretical uncertainties in the secondary flux in antiprotons, taking into
account the whole range of propagation parameters that is allowed by the existing B/C
data, again featured together with the existing low-energy data.
In the right panel of this figure, we have plotted $T_{\bar p}^3\Phi_{\bar p}$ in order to better illustrate the expected near $T_{\bar p}^{-3}$ scaling of the flux at high energies.
}
\label{fig_uncertainties}
\end{figure*}

%Discussion of Fig.~\ref{fig_contributions}
%

{
In Fig.~\ref{fig_contributions}, the various contributions to the
secondary antiproton flux, $\Phi = {v \, \psi}/{4 \pi}$ , from the spallation of interstellar H and He
by cosmic ray protons and alpha particles are presented
together with the existing low-energy data. Galactic propagation parameters
corresponds to the `medium' configuration of Table~\ref{tab_prop}, to be discussed later.
Solar modulation has been implemented through the force field approximation \cite{perko}, with a
Fisk potential $\phi_{\rm F}$ of 500 MV corresponding to the minimum of
 solar activity during which the observations have been performed.
Above an energy of a few tens of GeV, the antiproton flux decreases
like a power law.
If the scaling violations of the differential production cross section
${d \sigma}/{dE_{\pbar}}$ were negligible, the secondary source term
$q_{\pbar}^{\rm sec}(r , E)$ would have the same energy dependence as
the impinging cosmic ray flux
$\Phi = {v \, \psi}/{4 \pi} \propto E^{- \gamma}$. As discussed at the end
of section~\ref{sec_prop}, the antiproton propagator $G_{\pbar}$ scales
at high energy like $1/K \propto E^{- \delta}$. We expect the antiproton
flux to have a typical spectral behaviour like
$G_{\pbar} \times \Phi \propto E^{- \gamma - \delta}$.
The cosmic ray proton and helium fluxes have been borrowed from~\cite{don01}
where a fit of the BESS~\cite{sanuki00} and AMS~\cite{alcaraz00} data is
featured. The spectral index $\gamma$ is found to be equal to 2.72 for
protons and to 2.74 for helium. Would hadronic interactions be scale
invariant, the secondary antiproton flux of Fig.~\ref{fig_contributions}
-- for which $\delta = 0.7$ -- would drop like $\sim E^{-3.4}$. The actual
spectrum is slightly harder with an $E^{-3.3}$ energy dependence.
As a cross-check, we also took for comparison the parameterization of \cite{duperray03} (instead of \cite{Tan_Ng_82,Tan_Ng_83}) for the antiproton production cross section through ${\rm p \, H \to \bar{p}X}$. In this case we find qualitatively the same spectrum,
with a slightly harder spectral index of 3.2 instead of 3.3. This corresponds to an uncertainty in the background flux which is much smaller than the one induced by the not fully determined propagation parameters -- see the discussion below. 
}

\begin{table}[t!]
{
\begin{tabular}{|c||c|c|c|c|c|c|}
\hline
Case  & $\delta$ & $K_0$ [kpc$^2$/Myr] & $L$ [kpc] & $V_{C}$ [km/s] & $V_{a}$ [km/s] \\
\hline \hline
max  & 0.46 &  0.0765 & 15 &  5   & 117.6 \\
med  & 0.70 &  0.0112 & 4  & 12   &  52.9 \\
min  & 0.85 &  0.0016 & 1  & 13.5 &  22.4 \\
\hline
\end{tabular}}
\caption{Typical combinations of diffusion parameters that are compatible with the B/C
analysis \cite{mau01}; as demonstrated later, these parameter sets correspond to minimal, medium and
maximal primary antiproton fluxes, respectively.}
\label{tab_prop}
\end{table}

%Discussion of the range of propagation parameters allowed by B/C.
%

{
The semi-analytic treatment of cosmic ray propagation that is based on
the Bessel expansion~(\ref{bessel_psi}) is a convenient framework to
derive the theoretical uncertainties associated to the various parameters
at stake -- namely $K_{0}$, $\delta$, $V_{a}$, $V_{C}$ and the thickness $L$.
The space of these propagation parameters has been extensively scanned~\cite{mau01}
in order to select the allowed regions where the predictions on B/C -- a typical
cosmic ray secondary to primary ratio -- match the observations. Several
hundreds of different propagation models have survived that crucial test.
The propagation parameters are thus only loosely constrained by the cosmic ray nuclei abundances
so far observed.
The same conclusion has been reached independently in Ref.~\cite{Lionetto:2005jd}
with the help of a fully numerical code~\cite{Strong:1998pw} in which the
convective wind $V_{C}$ increases linearly with vertical height $z$.
However, the B/C ratio could not be accounted for when both galactic
convection and diffusive reacceleration were implemented at the same time,
a problem which our Bessel treatment does not encounter.
}

%Discussion of Fig.~\ref{fig_uncertainties} and of the yellow band.
%

{
The yellow band shown in Fig.~\ref{fig_uncertainties} is the envelope of the
secondary antiproton spectra computed with the set of $\sim 1,600$
different propagation models found in~\cite{mau01} to pass the B/C test.
This band comprises the theoretical uncertainty in the determination of
 the secondary
antiproton flux. It is confined by the `minimal' and `maximal'
configurations of Table~\ref{tab_prop}.
As a first observation, we notice how narrow the uncertainty strip is between $\sim$ 10 and 100 GeV.
The PAMELA and AMS-02 collaborations will thus be able to highlight even small spectral
deviations in that energy range. Above $\sim 100$ GeV, the yellow band
widens as a result of the energy dependence of the diffusion coefficient
$K$: from the B/C analysis, the spectral index $\delta$ may take any value
between $0.46$ and $0.85$; its spread $\Delta \delta = 0.4$ thus translates
into a factor of $10^{0.8} \sim 6$ of uncertainty on the antiproton flux
at 10 TeV -- two decades above the energy where the yellow strip is still
the thinnest. In fact, we see this expectation confirmed in the right 
panel of Fig.~\ref{fig_uncertainties}, where we can read off a ratio of 1 to 6 (at 10 TeV) between the minimal and maximal antiproton
flux expectations.
This large uncertainty in the secondary antiproton background at TeV
energies may look depressing.
One should keep in mind, however, that PAMELA and AMS-02 will considerably improve the
measurements of the cosmic ray nuclei abundances with a 
determination of the B/C ratio to a better accuracy and over a wider energy
 range than available so far.
This will translate into improved constraints on the propagation parameters
and eventually into a thinner uncertainty strip in the panels of
Fig.~\ref{fig_uncertainties}.
The antiproton spectrum itself will also be measured up to a few TeV in the
case of AMS-02. Once it is compared to the cosmic ray proton and helium
fluxes, the spectral index $\delta$ should be better determined.
Finally, we expect the LHC to improve the accuracy of the antiproton
production cross sections of the various nucleus-nucleus interactions
at stake.

When going to very high energies, a word of caution is nevertheless mandatory. The antiproton flux
rapidly decreases with energy and becomes vanishingly small above 1 TeV.
Even with the AMS-02 large acceptance, the statistical error on a
measurement of the antiproton background at $\sim 800$ GeV is comparable
to the present theoretical uncertainty as featured in the left panel of
Fig.~\ref{fig_comp}: the vertical error bars of the highest energy data
point extend over the yellow uncertainty strip. In fact, above 10 TeV we expect not more than $\mathcal{O}(1)$ antiprotons per $m^2$ and \emph{year}, 
posing a formidable (if surmountable) challenge even for follow-up experiments. In this article, we therefore restrict our analysis throughout to an energy range $\lesssim10\,$TeV.

}

%%%%%%%%%%%%%%%%%%%%%%%%%%%%%%%%%%%%%%%%%%%%%%%%%%%%%%%%%%%%%%%%%%%%%%%%%%%%%%
%%%%%%%%%%%%%%%%%%%%%%%%%%%%%%%%%%%%%%%%%%%%%%%%%%%%%%%%%%%%%%%%%%%%%%%%%%%%%%
\section{Antiprotons from dark matter annihilations in the galactic halo}
\label{sec_signal}

The source function $q_{\pbar}^{\rm prim}(T , \mathbf r)$ describes the number of primary antiprotons per unit time, energy and
volume element that are produced with a kinetic energy $T$ at a given position $\mathbf
r$ in the galactic halo:
\be
q_{\pbar}^{\rm prim}(T , \mathbf r) =
\frac{1}{2}  \left<\sigma_\mathrm{ann}v\right> \,
\left\{ \frac{\rho_{CDM}(\mathbf r)}{m} \right\}^2 \;
\sum_f \, B^f \, \frac{\mathrm{d}N^f}{\mathrm{d}T} \,.
\label{source}
\ee
Here, $\left<\sigma_\mathrm{ann} v\right>$ is the DM self-annihilation rate, $\rho_{CDM}$
the DM halo density and $m$ the DM particle's mass. The sum runs over all annihilation
channels $f$, where $B^f$ are the branching ratios
and $\mathrm{d}N^f/\mathrm{d}T$ the fragmentation functions into antiprotons,
respectively. The factor of 1/2 has to be included whenever the DM particle is its own
antiparticle and therefore only annihilates in pairs (which is the case for the DM
candidates that we will consider below).

The source function can thus be disentangled into an astrophysical part, given by the
dark matter distribution $\rho_{CDM}$, and a particle physics part that comprises the
remaining information. The first one is usually subject to much greater uncertainties
than the latter and will be dealt with in the following subsection. In section
\ref{sec_models}, we will then address the particle physics part by discussing three
benchmark scenarios with DM masses in the TeV range, with their respective expressions
for the annihilation cross section and branching ratios. The fragmentation functions
$\mathrm{d}N^f/\mathrm{d}T$, finally, can be obtained from Monte Carlo programs like
\textsc{Pythia} \cite{pythia}. We use here the tabulated fragmentation functions of the
\textsc{DarkSUSY} package \cite{ds}, which are based on a large number of \textsc{Pythia}
runs ($10^6$ per annihilation channel and mass).

\subsection{The distribution of dark matter in the Milky Way}

\subsubsection{Halo profiles}
\label{sec_profiles}

\begin{table}[t]
    {
   \begin{tabular}{|l||c|c|c||c|c|c|}
        \hline
    \textbf{Halo model}  & $\alpha$ & $\beta$ & $\gamma$ & $\rho_\mathrm{s}$ [$10^6
M_\odot \mathrm{kpc}^{-3}$] & $r_\mathrm{s}$ [$\mathrm{kpc}$]\\
        \hline \hline
       isothermal sphere & 2   & 2 & 0    & 7.90 & 4     \\
       NFW 97 \cite{Navarro:1996gj}  & 1   & 3 & 1    & 5.38 & 21.75 \\
       Moore 04   \cite{Moore:1999nt}  & 1   & 3 & 1.16 & 2.54 & 32.62 \\
       Moore 99    \cite{Diemand:2004wh} & 1.5 & 3 & 1.5  & 1.06 & 34.52 \\
        \hline
   \end{tabular}}
  \caption{Parameters in Eq.~(\ref{prof}) for the halo models that we consider here.
Scale radius $r_\mathrm{s}$ and density $\rho_\mathrm{s}$ are strongly correlated with
the virial mass of the galaxy \cite{Eke:2000av} and we adopt the values obtained in \cite{for04} for the Milky Way.
In the case of the Moore 99 profile, DM self-annihilations set an upper bound to the
maximal possible density and we follow the usual prescription of imposing a cutoff radius
inside which $\rho_{CDM}$ is assumed to be constant \cite{ber92}.
When the DM distribution is cuspy -- for $\gamma \geq 1$ -- we smooth it so as to
keep the total number of annihilations constant -- see the text for further details.
\label{tab_halo}}
\end{table}

The Milky Way halo distribution is only poorly constrained by direct observations and its
form must therefore be inferred from N-body simulations of gravitational clustering. Due
to the restricted resolution of these simulations, however, the innermost slope of the
density profile can only be obtained by an extrapolation of the behaviour at larger radii
(i.e. $r\gtrsim0.1$ kpc) and is thus bound to a considerable amount of uncertainty;
furthermore, it is still an (numerically) unsolved issue how to correctly include the
effect of baryons during the gravitational collapse.
To account for this situation, we will here for illustrative purposes restrict ourselves
to a choice of four different profiles that basically span the whole range of reasonable
halo models with respect to indirect dark matter detection prospects; the isothermal
sphere, the Navarro-Frenk-White (NFW) \cite{Navarro:1996gj} and the Moore profiles \cite{Moore:1999nt,Diemand:2004wh}. Like most of their
relatives, these halo models can be parameterized as
\be
\label{prof}
\rho_\mathrm{CDM}(\mathbf{r}) = \rho_\mathrm{s} \,
\left( { \frac{r_\mathrm{s}}{r}} \right)^{ \gamma} \,
\left\{ 1 + \left( \frac{r}{r_\mathrm{s}} \right)^{ \alpha}
\right\}^{ \frac{\gamma - \beta}{\alpha}} \;\; ,
\ee
where the corresponding parameters are summarized in Table \ref{tab_halo}. Note that at
the sun's distance to the galactic center, $r=R_0\equiv8.5$ kpc, one recovers for all
profiles a local halo density of $\rho_0=\rho_\mathrm{CDM}(R_0)\sim0.3$ GeV cm$^{-3}$.

%Smoothing the central cusp.
%

{
Our semi-analytic treatment of cosmic ray galactic propagation relies
on Bessel expansions like~(\ref{bessel_psi}). The function
$J_{0} \left( \alpha_{i} \, r / R \right)$ probes details as small
as $\lambda \sim 2 \pi R / \alpha_{i}$ where $\alpha_{i}$ is the i-th
zero of the function $J_{0}$ and $R$ denotes the radius of the diffusive
halo. In the case of an NFW or Moore profile, the correct description
of the central cusp -- where the annihilation rate diverges like
$r^{- 2 \gamma}$ -- would imply the necessity of an infinite series of such Bessel
functions. This would lead to unacceptable CPU time and to a lack of
numerical accuracy. Following the method explained in~\cite{bar05},
we have renormalized the DM central distribution without modifying
the absolute number of DM particle annihilations.
Actually, within a small sphere of radius $r_{c}$, we have replaced
the density profile~(\ref{prof}) which diverges like
${\rho(r)}/{\rho_{c}} = ({r_{c}}/{r})^{\gamma}$ where
$\rho_{c} \equiv \rho(r_{c})$, by the milder distribution
\be
\left\{
{\displaystyle \frac{\rho(r \leq r_{c})}{\rho_{c}}} \right\}^{2}
= 1 + \left\{
{\displaystyle \frac{2 \, \pi^{2}}{3}} \,
\left( \xi - 1 \right) \,
\sin_{c}^{2} \left( {\displaystyle \frac{\pi \, r}{r_{c}}} \right)
\right\}\,,
\label{profile_convergent}
\ee
where $\sin_{c}(x) \equiv \sin(x) / x$. That renormalized density
leads to the same number of DM annihilations as the actual cusp
provided that
\be
\xi \; = \;
{\displaystyle \frac{3}{3 \, - \, 2 \, \gamma}} \;\; ,
\ee
in the case of NFW 97 and Moore 04. Note that for the Moore 99 cusp, the
actual annihilation rate diverges logarithmically and we have to impose
a cutoff on the DM density, $\rho_\mathrm{CDM}\lesssim\rho_{\rm ann}$. 
This upper bound 
arises from the increasing DM self-annihilation rate in regions of enhanced DM densities and is given by:
\beq
\rho_{\rm ann} \; = \;
{\displaystyle \frac{m}{\left<\sigma_\mathrm{ann}v\right> \, \tau}}
\, ,
\label{cutoff_CDM_density}
\eeq
where an age of $\tau = 12$ Gy has been assumed for the central
DM concentration. This cutoff translates into a factor of
\beq
\xi_{\rm \, Moore \, 99} \; = \; 2 \, \ln
\left( {\displaystyle \frac{\rho_{\rm ann}}{\rho_{c}}} \right) \;\; .
\eeq
In practice, we have set $r_{c} = 500$ pc and pushed the Bessel
expansion up to the 200-th order. We have accelerated its convergence
by using the method explained in~\cite{bringmann_phd}.
Because the antiproton propagator $G_{\pbar}$ that connects the
solar system to the galactic center varies smoothly over that region,
a smaller core radius $r_{c}$ would not appreciably change our results.
}

\begin{figure*}[t!]
  \begin{minipage}[t]{0.49\textwidth}
      \centering
   \includegraphics[width=\textwidth]{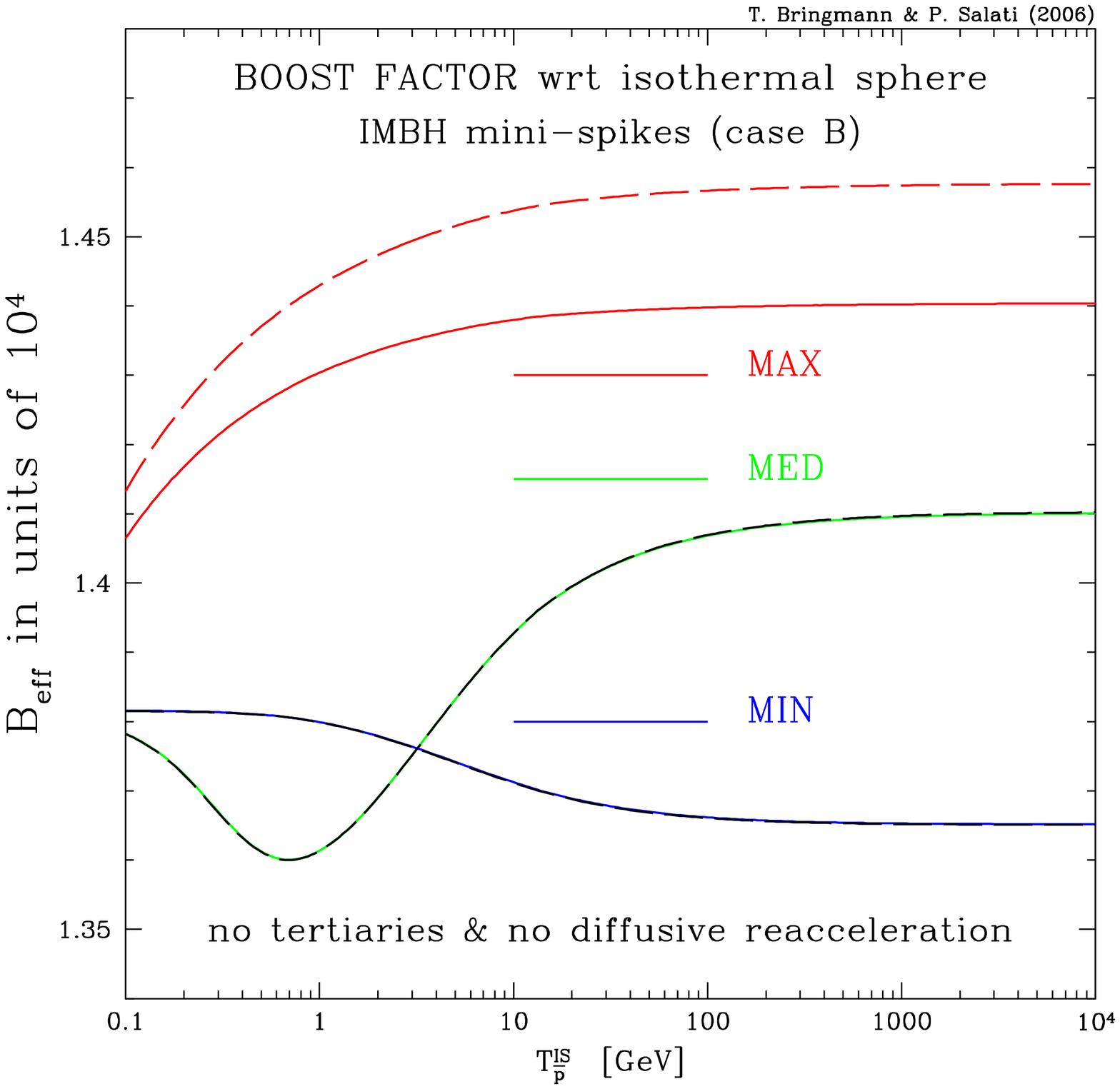}\\
   \end{minipage}
  \begin{minipage}[t]{0.02\textwidth}
   \end{minipage}
  \begin{minipage}[t]{0.49\textwidth}
      \centering   
  \includegraphics[width=\textwidth]{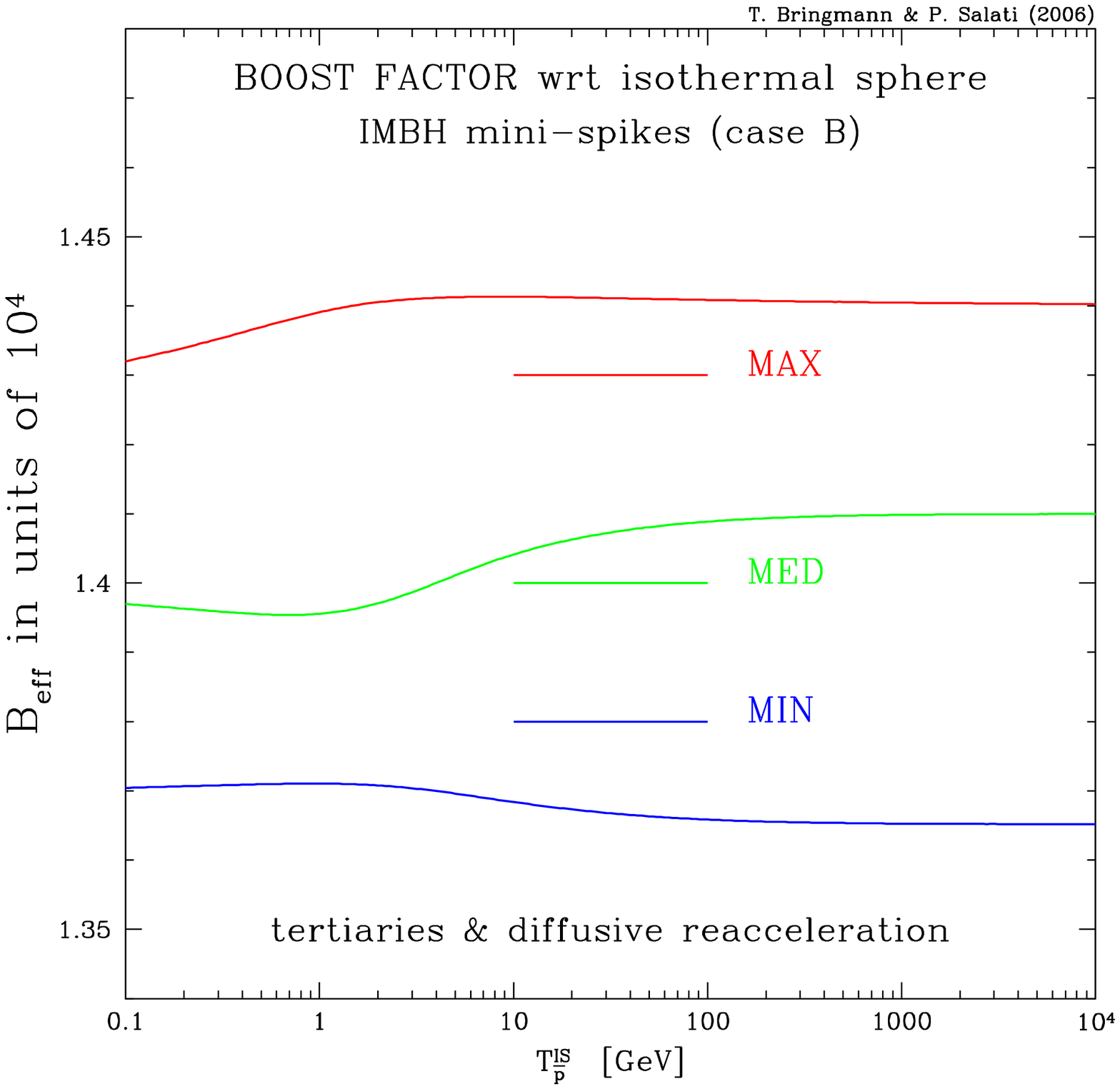}\\
   \end{minipage}  
  \begin{minipage}[t]{0.49\textwidth}
      \centering
  \includegraphics[width=\textwidth]{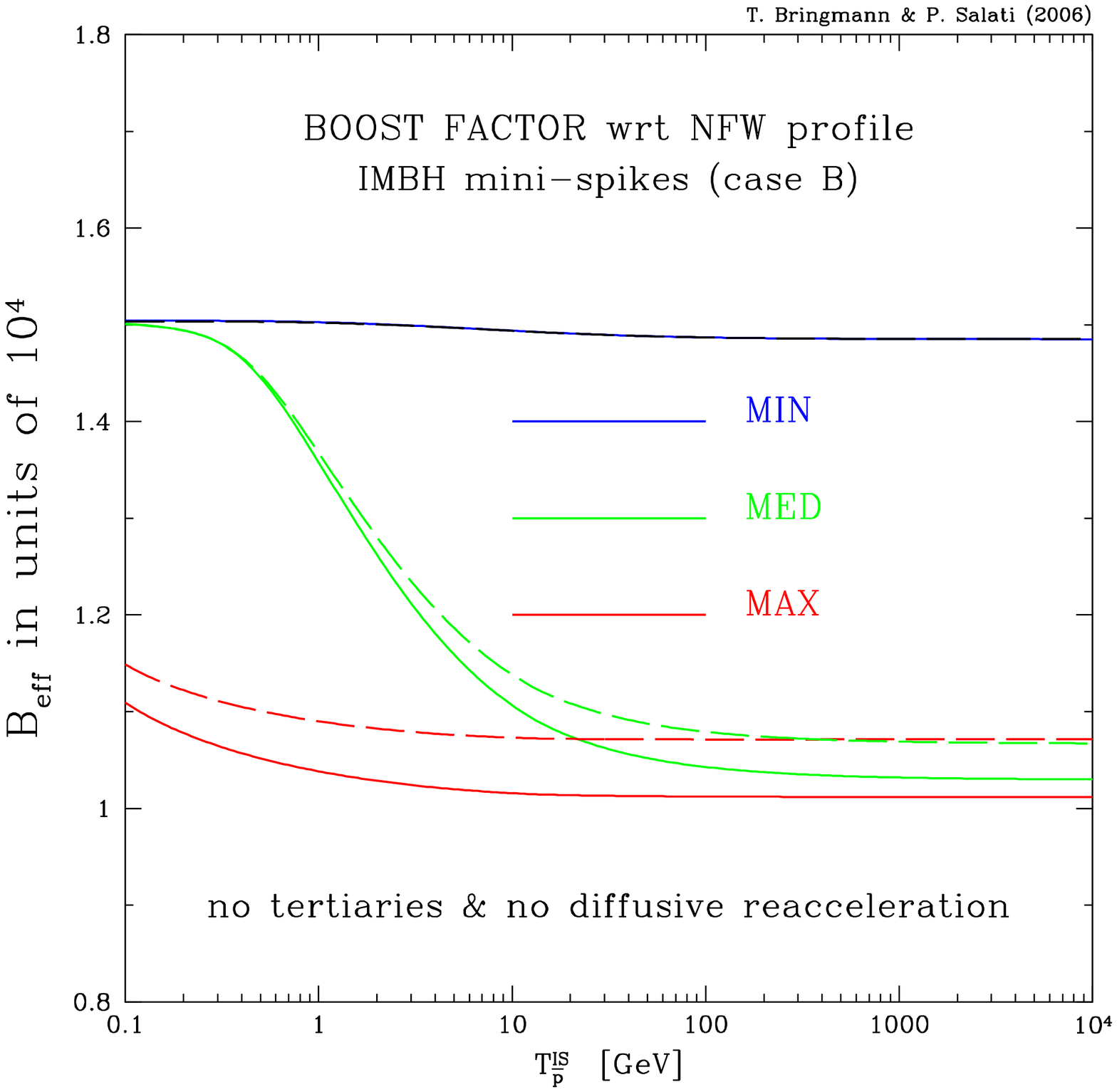}\\
   \end{minipage}
  \begin{minipage}[t]{0.02\textwidth}
   \end{minipage}
  \begin{minipage}[t]{0.49\textwidth}
      \centering   
  \includegraphics[width=\textwidth]{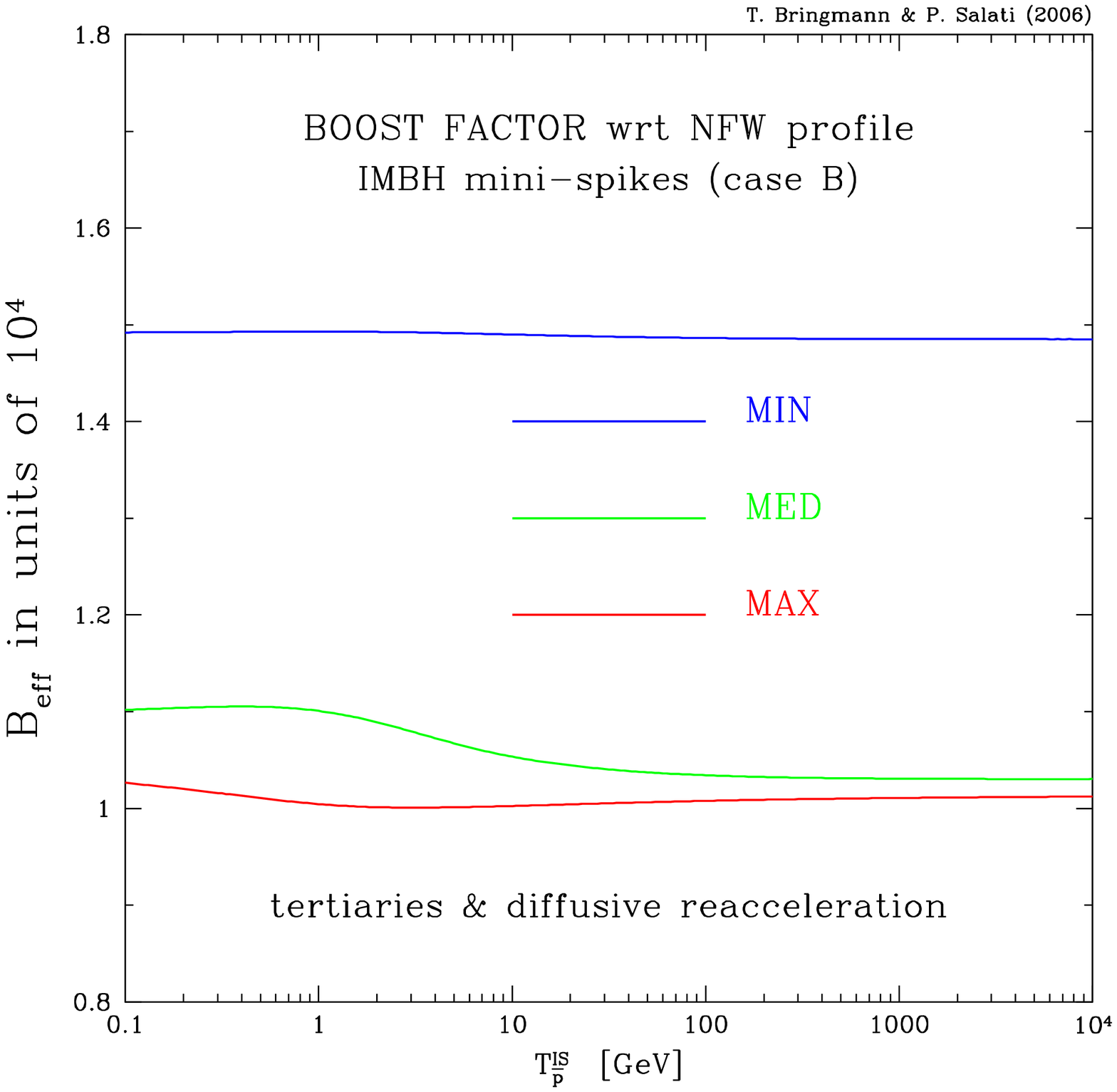}\\
   \end{minipage}
{
\caption{Boost factor for an average Milky Way halo population of IMBHs as a function of the antiproton kinetic energy.
The two panels on the left do not include the contribution from tertiaries, energy
losses and diffusive reacceleration. The long dashed lines have been obtained by
integrating directly the antiproton Green function $G_{\pbar}$ over the diffusive
halo.
In the right panels, tertiaries, energy losses and diffusive reacceleration have
been taken into account. An isothermal sphere -- top panels -- as well as an NFW
profile -- bottom panels -- have been considered for the DM smooth distribution.}
\label{fig_boost_propagator}}
\end{figure*}

\subsubsection{Boost factors}

Following the paradigm of hierarchical structure formation, it is very likely that the DM
distribution exhibits substructures (``clumps''), i.e. small inhomogeneities superimposed
on the smooth background profiles described above. On very small scales, e.g., one
expects a cutoff in the cold dark matter power spectrum \cite{gre05} and a considerable
number of the first gravitationally bound objects, with a mass around this cutoff, might
have survived until today \cite{die05}. Another interesting proposal is the existence of
a population of intermediate mass black holes (IMBHs) in our galaxy, each of them being
surrounded by a DM mini-halo with a rather steep profile, i.e.~a ``mini-cusp'' \cite{imbh}. 
The effect of halo substructures is in any case to enhance the flux of antiprotons -- or
any other primarily produced particles -- at earth, as compared to the case of a smooth
DM matter distribution. Such an enhancement immediately follows from the fact that
$\langle\rho^2\rangle$ is generally larger than $\langle\rho\rangle^2$, with the
difference increasing with the degree of inhomogeneity, and is often encoded in a
universal boost factor $b$. However, as has been stressed in \cite{lav06}, this
boost factor is actually not so universal in that it -- given a DM distribution
$\rho_{CDM}$ -- depends on both the particle species and the energy range under
consideration (see, e.g., \cite{bri05} for a similar observation).

As an example, and as a practical application of the formalism developed in \cite{lav06},
let us now determine the boost factor for antiprotons in the ``type B'' IMBH scenario of
\cite{imbh} -- which is representative of a class of models in which black holes originate
from massive objects formed directly during the collapse of primordial gas in early-forming 
halos. A complete discussion of the antiproton signal and of its statistical properties
is not in the focus of this paper and the reader is rather referred to the thorough analysis
of Ref.~\cite{brun06} for details. As a sneak preview of that work, we derive here
the effective boost factor
\beq
B_{\rm eff} \; = \; {\displaystyle
\frac{< \Phi_{\pbar \, , \, {\rm IMBH}} >}{\Phi_{\pbar \, , \, {\rm smooth}}}} \;\; ,
\eeq
for an average IMBH population of ``type B''. A smooth halo density $\rho_{\mathrm{CDM}}$
yields a flux
\beq
\Phi_{\pbar \, , \, {\rm smooth}} \propto
{\displaystyle \int} \, d^{3}{\mathbf x}_{S} \;\;
G_{\pbar}(M \leftarrow S , E) \,
\left\{
{\displaystyle \frac{\rho_\mathrm{CDM}(\mathbf{r})}{\rho_{0}}} \right\}^{2} ,
\eeq
whereas the IMBH contribution is given by
\beq
< \! \Phi_{\pbar \, , \, {\rm IMBH}} \! > \propto \,
{\displaystyle \int} \, d^{3}{\mathbf x}_{S} \;\;
G_{\pbar}(M \leftarrow S , E) \;
\xi_{\rm \, sp} \, n_{\rm \, sp}({\mathbf r}) \; .
\eeq
In Ref.~\cite{brun06}, the average number $n_{\rm \, sp}$ of mini-spikes per unit
volume has been fitted with the results of $\sim$ 200 different Monte-Carlo
realizations -- each containing eventually a hundred objects -- obtained in
Ref.~\cite{imbh} by evolving an initial distribution of IMBHs orbiting in the galactic
halo and by taking into account close encounters and their associated tidal disruptions.
The enhancement of the DM annihilation rate inside each mini-spike is described by the
source boost factor
\beq
\xi_{\rm \, sp} \; = \;
{\displaystyle \int_{\rm DM \, cloud}}
\left\{ {\displaystyle \frac{\delta \rho_{\rm CDM}({\mathbf r})}{\rho_{0}}} \right\}^{2}
\, d{\mathbf r} \;\; ,
\eeq
where $\rho_{0}$ is some value of reference generally set equal to
$\rho_\mathrm{CDM}(R_0)$. In a region of size $R_{\rm sp}$ around the IMBH, the DM
density is described by a power law $r^{-7/3}$. The annihilation rate diverges at the
center and we impose a cutoff on the DM density
$\rho_\mathrm{CDM} \lesssim \rho_{\rm ann}$ -- see Eq.~(\ref{cutoff_CDM_density}) and
Ref.~\cite{ber92}. The effective volume $\xi_{\rm \, sp}$ depends on the DM particle
mass $m$ and self-annihilation cross section $\left<\sigma_\mathrm{ann}v\right>$.
We have used here an average spike radius of $R_{\rm sp} = 2.84$ pc and an average
surface density of
$\delta \rho_{\rm CDM}(R_{\rm sp}) = 48.51 \; {\rm M_{\odot} \, pc^{-3}}$ -- see
once again Ref.~\cite{brun06} for details.

In the upper panels of
Fig.~\ref{fig_boost_propagator}, the isothermal sphere of Table~\ref{tab_halo}
corresponds to the smooth distribution of reference with respect to which the
enhancement of the antiproton signal is computed.
A DM particle mass $m = 1$ TeV and a self-annihilation cross section
$\left<\sigma_\mathrm{ann}v\right> = 3 \times 10^{-24}$ cm$^{3}$ s$^{-1}$ lead
to a mini-spike effective volume of
$\xi_{\rm \, sp} = 1.2 \times 10^{5}$ kpc$^{3}$. In the lower panels where
the isothermal sphere is replaced by an NFW profile, a DM particle mass of 1 TeV
and a self-annihilation cross section of $3 \times 10^{-26}$ cm$^{3}$ s$^{-1}$
have been assumed, hence a value of
$\xi_{\rm \, sp} = 3.1 \times 10^{6}$ kpc$^{3}$.
%
%sigma_v_an = 3.0e-26;//[cm^{3} s^{-1}] cas 3_26
%XSI_CLUMP_IMBH = 3.12747e+06 [kpc^{3}]
%
%sigma_v_an = 300.0e-26;//[cm^{3} s^{-1}] cas 3_24
%XSI_CLUMP_IMBH = 1.16573e+05 [kpc^{3}]
%
In the left panels, cosmic ray propagation does not include the contribution from
tertiaries, nor energy losses and diffusive reacceleration. Switching off these
processes allows to perform the calculation of $B_{\rm eff}$ by the two methods
presented in section~\ref{sec_prop}. The convolution over the diffusive halo of
the antiproton propagator $G_{\pbar}$ with the square
$\{ {\rho_\mathrm{CDM}(\mathbf{r})}/ {\rho_{0}} \}^{2}$ of the smooth DM density
profile~(\ref{prof}) and with the IMBH source term
$\xi_{\rm \, sp} \, n_{\rm \, sp}({\mathbf r})$ leads to the long-dashed curves.
Alternatively, we have directly included the IMBH source term
$\xi_{\rm \, sp} \, n_{\rm \, sp}({\mathbf r})$ in our Bessel code to generate
the solid lines. The injection spectrum of antiprotons cancels out in the
calculation.
The agreement between these two different approaches is astonishingly good.
The discrepancies do not exceed a few percent in the worst case. This occurs
in particular for `maximal' propagation where the range of the antiproton
propagation is large and allows to probe the radial boundaries of the diffusive
halo. In the Bessel approach, the cosmic ray density vanishes at $r = R$.
This condition is not implemented in the propagator approach
even if the contribution of the outward regions is not taken into account.
This leads to an overestimation of $B_{\rm eff}$ as is particularly clear
in the upper-left pannel.
At first glance, the boost factor depends very weakly on the propagation model.
As the antiproton energy increases, so does the propagation range. In the case
of the isothermal sphere, the population of galactic IMBHs -- whose number
density is given on average by  $n_{\rm \, sp}({\mathbf r})$ -- tends to
contribute more to the signal than the smooth distribution. As the propagation
range increases, more and more mini-spikes come into play. This
results into the increase of $B_{\rm eff}$ in the upper-left pannel for the
`medium' and `maximal' propagation configurations.
On the contrary, in the case of an NFW profile, as soon as the propagation
range reaches the galactic center and its strong DM annihilation site, the
boost factor drops as is particularly clear for the `medium' configuration of
the lower-left pannel.
In the right panels, tertiary production, energy losses and diffusive
reacceleration have been switched on with the consequence of smoothing
the variations of the boost factor with energy.
In Fig.~\ref{fig_boost_propagator}, the values of $\xi_{\rm \, sp}$ have
been adjusted in order to get a boost of $\sim 10^{4}$ -- though the DM
particle mass and self-annihilation cross section are reasonable. Because
the distribution of mini-spikes inside the galactic halo is not unique,
the boost factor $B_{\rm eff}$ is subject to large variations -- typically
a factor of a few exceeding 10 at low antiproton energies -- depending on
the actual IMBH distribution. For a full analysis of the associated boost
uncertainty, we refer to~\cite{brun06} -- but would like to mention already
here that we expect this variance to decrease significantly at the high energies
considered here.

\subsection{Dark matter masses at the TeV scale}
\label{sec_models}

Searching for possible primary contributions to the antiproton spectrum at very high energies,
we have to focus on the annihilation of rather heavy (i.e.~at least TeV scale) DM particles.
While such high masses can appear in a variety of models, we will restrict ourselves in
the following to the most popular and most commonly studied scenarios, i.e.~supersymmetry
and extra dimensions. In this section, we introduce the DM candidates that arise in these
scenarios and in each case briefly describe the main features that are relevant in our
context.  First, however, note that the usual WIMP relation,
\be
  \label{wimpomega}
  \Omega_\mathrm{WIMP}h^2\sim
\frac{3\cdot10^{-27}\,\mathrm{cm}^3\,\mathrm{s}^{-1}}{\langle\sigma v\rangle}\,,
\ee
in general rather prefers slightly smaller masses. To balance the reduced annihilation
cross sections one expects for higher masses, one has thus to rely on the presence of
effective coannihilation channels with other particles during the freeze-out process. In
accordance with this observation, the mass splitting between the DM particle and the
next-to-lightest non-standard model particles is generically very small in all the
examples described below.

Let us now start with the case of supersymmetry, where the lightest stable supersymmetric
particle (LSP) provides an excellent dark matter candidate \cite{WIMPDM}. In most models,
it is given by the (lightest) neutralino, which is a linear combination of the
superpartners of the gauge and Higgs fields,
\be
  \chi\equiv\tilde\chi^0_1= N_{11}\tilde B+N_{12}\tilde W^3 +N_{13}\tilde
H_1^0+N_{14}\tilde H_2^0\,.
\ee
While the neutralino is often a gaugino, with a large Bino fraction and a mass of a
couple of hundred GeV or less, the hyperbolic branch/focus point region of minimal
supergravity (mSUGRA) typically exhibits very heavy neutralinos with a large Higgsino
fraction \cite{Higgsino}. From the requirement that Higgsinos should give the right relic
density, their mass has to be around 1 TeV \cite{pro04}. For these high masses, the
neutralino is an almost pure (anti-) symmetric combination of the two neutral Higgsino
states  -- in which case the annihilation cross section into $Z$ (as well as into Higgs
boson) pairs vanishes exactly and  that into quarks is usually heavily suppressed by
multi-TeV squark masses in the propagator. For these reasons the annihilation into $W$
bosons typically clearly dominates; depending on the actual neutralino composition,
however, a significant fraction can also go into $Z$ bosons or heavy quark pairs.

Another interesting situation arises in the case of a neutralino that is almost a pure
Wino, as expected for example in anomaly mediated supersymmetry breaking (AMSB) scenarios
\cite{ull01}. For Winos, the preferred mass from relic density requirements is peaked at
about 1.7 TeV \cite{pro04} (a \emph{pure} Wino state obtains the right relic density for $m\sim2.2\,$TeV; {taking into account non-perturbative effects, this mass is further increased by about $600\,$GeV \cite{Hisano:2006nn})}. Non-perturbative, binding energy effects then result in
greatly enhanced annihilation cross-sections today, when the neutralinos have very small
galactic velocities \cite{his04}. In this limit, heavy Winos annihilate almost
exclusively into gauge bosons (as a side remark, the annihilation into photons is also
significantly enhanced w.r.t. the non-perturbative result, leading to promising prospects
for an indirect detection in terms of gamma rays \cite{his04,ber05,Chattopadhyay:2006xb}).

\begin{table*}[ht!]
   \begin{tabular}{|l||c|c|c|c|c|c|c|c|c|c|c|c|}
        \hline
    \textbf{DM model}  & $m$ &  $\left<\sigma_\mathrm{ann} v\right>$ & $t\bar t$ & $b\bar
b$ & $c\bar c$ & $s\bar s$ & $u\bar u$ & $d\bar d$ & $ZZ$ & $W^+W^-$ & $HH$ & $gg$\\
        \hline \hline
       LSP1.0 & 1.0 & 0.46 & - & - & - & - & - & - & - & 100 & - & -     \\
       LKP1.0 & 1.0 & 1.60 & 10.9 & 0.7 & 11.1 & 0.7 & 11.1 & 0.7 & 0.5 & 1.0 & 0.5 & 0.5
\\
       LSP1.7 & 1.7 & 102 & - & - & - & - & - & - & 20.1 & 79.9 & - & -     \\
       LKP1.7 & 1.7 &  0.55 & 11.0 & 0.7 & 11.1 & 0.7 & 11.1 & 0.7 & 0.5 & 0.9 & 0.5 &
0.5   \\
        \hline
   \end{tabular}
  \caption{Our benchmark models for studying possible primary contributions to the
antiproton spectrum at high energies; $m$ is the DM particle's mass (in TeV),
$\left<\sigma_\mathrm{ann} v\right>$ its annihilation rate (in $10^{-26}$cm$^3$s$^{-1}$)
and the remaining columns give the branching ratios into the annihilation channels
relevant for $\bar p$ production (in percent). The corresponding values are typical for
high Higgsino (LSP1.0) and Wino (LSP1.7) fractions of the neutralino; for the latter we
take the non-perturbative expressions from \cite{his04}, while for the former we have
calculated the annihilation cross section of a pure (anti-)symmetric Higgsino into $W$
bosons and choose to neglect other annihilation channels (see text for further details). In the
case of the LKP, the quoted values for the branching ratios, as well as
$\left<\sigma_\mathrm{ann} v\right>m^2$, are actually very insensitive to the parameters of the
model (see, e.g., \cite{bri05}).}
  \label{tab_models}
\end{table*}

The third example of a TeV scale dark matter candidate we want to consider here is that
of the lightest Kaluza-Klein particle (LKP) in models with universal extra dimensions
(UED) \cite{ued}, where all standard model fields are allowed to propagate in a
higher-dimensional bulk. After compactification of the internal space, these additional
degrees of freedom appear as towers of new, heavy states in the effective
four-dimensional theory; the stability of the lightest of these states, i.e.~the LKP, is
guaranteed by the existence of an internal ${Z}_2$ symmetry (called KK parity) that
derives from higher-dimensional translational invariance. Taking into account radiative
corrections to the KK masses, the LKP is expected to be well approximated by the $\B$,
the first KK excitation of the weak hypercharge boson \cite{LKP}. Detailed relic density
calculations show that it can account for the required dark matter density if the
compactification scale (and thus the $\B$ mass) lies in the range $0.6\lesssim
m_\B\lesssim1.4$ TeV \cite{kak06}, mainly depending on the standard model Higgs mass;
deviations from the minimal scheme for calculating the radiative mass spectrum weaken the
upper bound on the compactification scale to about 2 or 3 TeV \cite{kon05}. The main
annihilation channels of the $\B$ that are relevant for antiproton production are those
into quark pairs (about 35 \% in total); the annihilation into gauge and Higgs bosons is
of the order of \mbox{1 \%} each and thus subdominant.

Finally, let us stress once more that in all these examples of heavy WIMP candidates, TeV
masses appear in a very natural way. Furthermore, one may easily find even higher masses
for e.g.~neutralino dark matter as soon as one leaves minimal prescriptions for
supersymmetry breaking \cite{pro05}. However, since the annihilation cross section in
general scales as $\left<\sigma_\mathrm{ann} v\right>\propto m^{-2}$, the number of
primary antiprotons is suppressed as
\be
q_{\pbar}^{\rm prim} \propto m^{-4} \, .
%Q\propto m^{-4}\,.
\ee
This means that even if the background flux scales roughly as $T^{-3}$ (see Section
\ref{sec_background}), it is very unlikely that one will be able to discriminate a
primary antiproton contribution from DM particles with masses significantly higher than 1
TeV -- unless one allows for artificially high boost factors \footnote{
Note that this argument can be evaded if one encounters resonant DM annihilations that
still -- as in the case of heavy Winos \cite{his04} -- do not significantly alter the
relic abundance.
}. Since also the total rates become very low far beyond 1 TeV, we do not consider such
high DM masses in the following.

\subsection{Future detectability}

After having introduced typical scenarios with high DM masses, let us now choose as
benchmark models a 1 TeV pure Higgsino, a 1.7 TeV pure Wino and a $\B$ with corresponding
masses (having in mind, of course, a direct comparison between the LKP and the LSP). In
Table~\ref{tab_models}, we summarize the main properties of these DM candidates.
Before we now proceed to study possible imprints in the antiproton spectrum for our
benchmark models, we note that the cases of a pure Wino \cite{his05} and the LKP
\cite{bri05,bar05} have already attracted some attention before.

\begin{figure*}[t]
  \begin{minipage}[t]{0.49\textwidth}
      \centering
  \includegraphics[width=\textwidth]{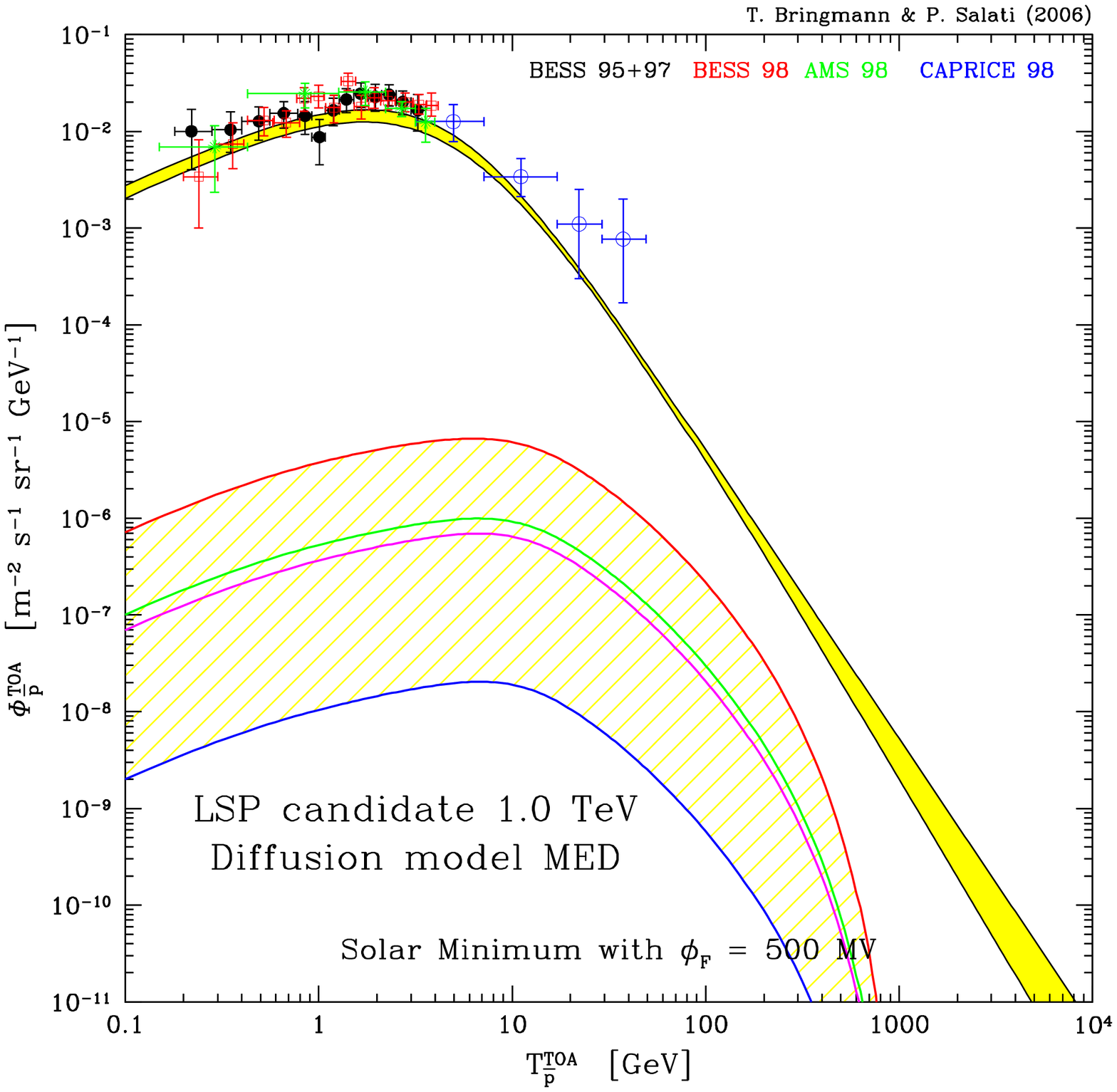}\\
   \end{minipage}
  \begin{minipage}[t]{0.02\textwidth}
   \end{minipage}
  \begin{minipage}[t]{0.49\textwidth}
      \centering   
  \includegraphics[width=\textwidth]{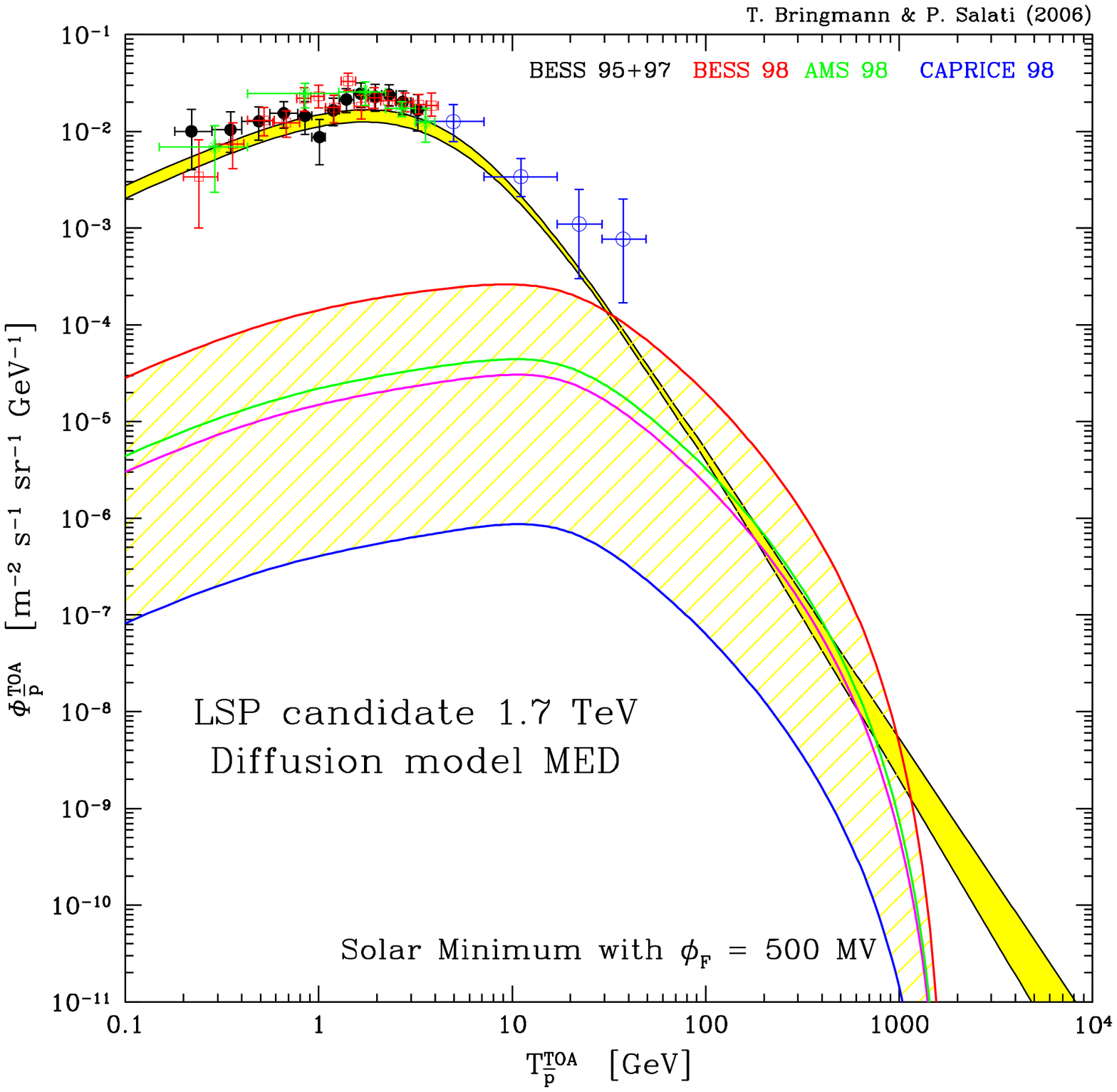}\\
   \end{minipage}  
  \begin{minipage}[t]{0.49\textwidth}
      \centering
  \includegraphics[width=\textwidth]{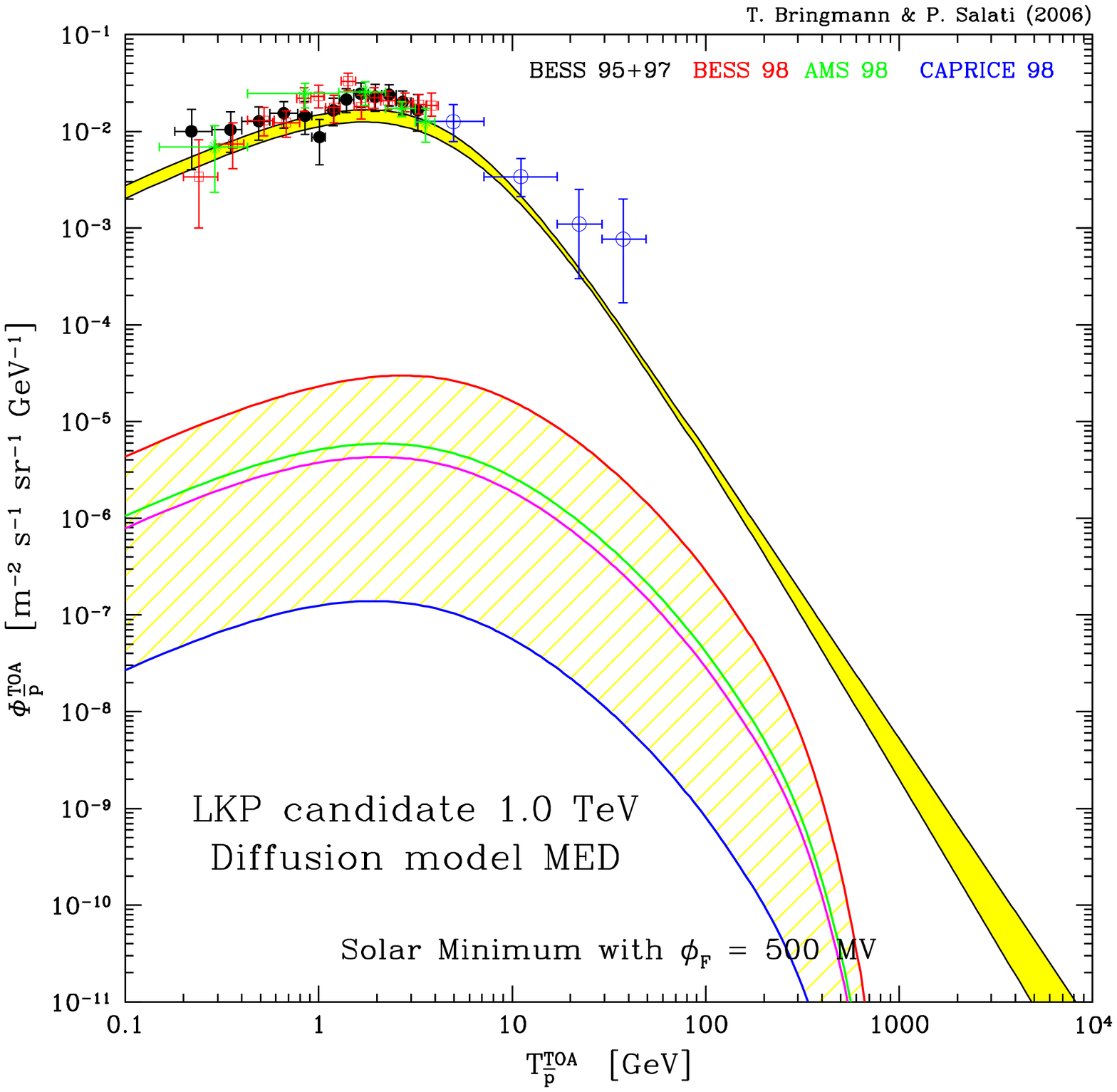}\\
   \end{minipage}
  \begin{minipage}[t]{0.02\textwidth}
   \end{minipage}
  \begin{minipage}[t]{0.49\textwidth}
      \centering   
  \includegraphics[width=\textwidth]{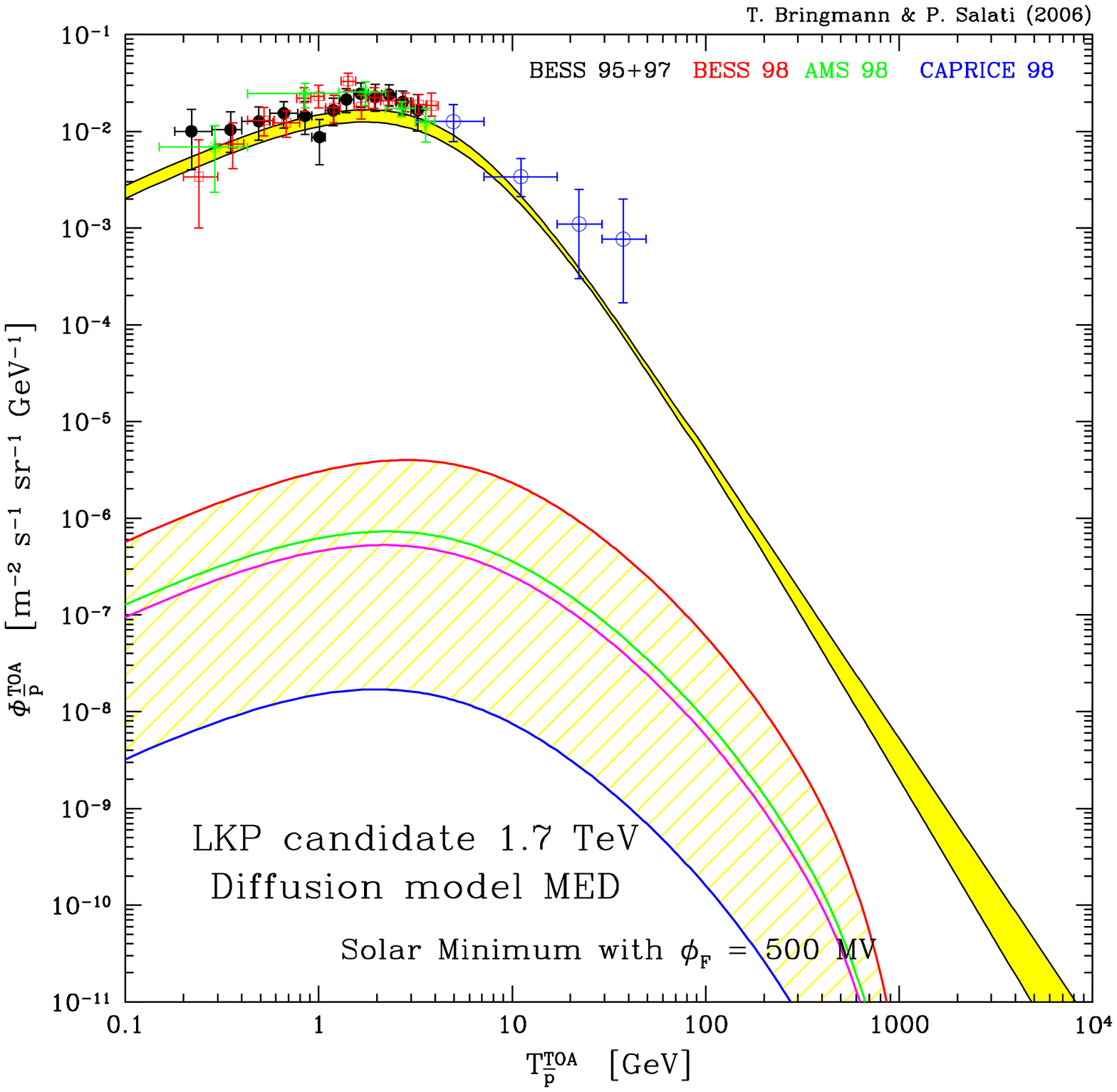}\\
   \end{minipage}
 \caption{The primary flux in antiprotons, as compared to the background in secondaries,
for the set of benchmark models specified in Table \ref{tab_models}. From bottom to top,
the different curves correspond to the isothermal sphere, NFW, Moore 04 and Moore 99 profiles,
respectively; for the diffusion parameters we adopt the `medium' configuration of Table
\ref{tab_prop}.}
 \label{fig_prof}
\end{figure*}

\begin{figure*}[t]
  \begin{minipage}[t]{0.49\textwidth}
      \centering
  \includegraphics[width=\textwidth]{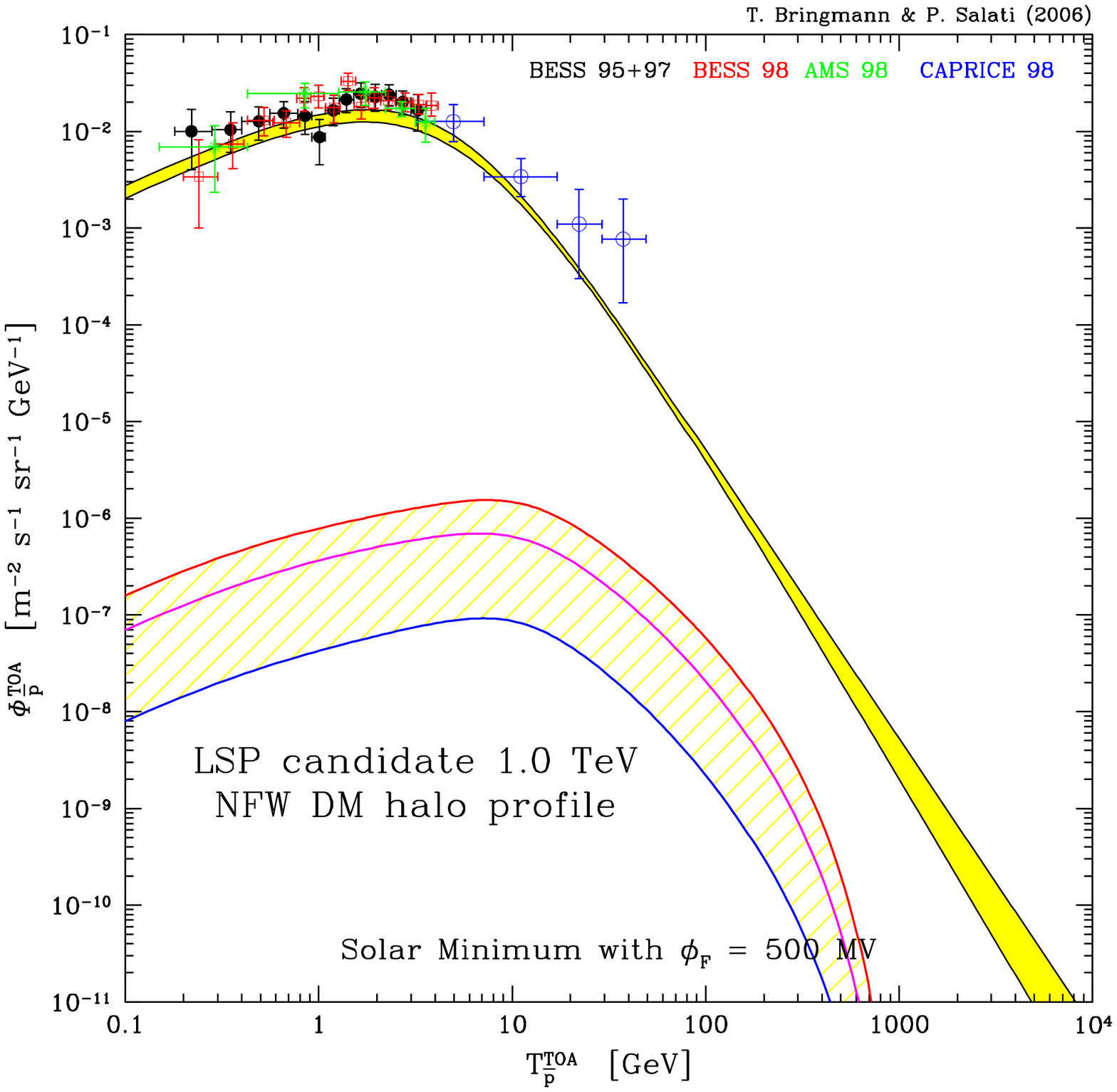}\\
   \end{minipage}
  \begin{minipage}[t]{0.02\textwidth}
   \end{minipage}
  \begin{minipage}[t]{0.49\textwidth}
      \centering   
  \includegraphics[width=\textwidth]{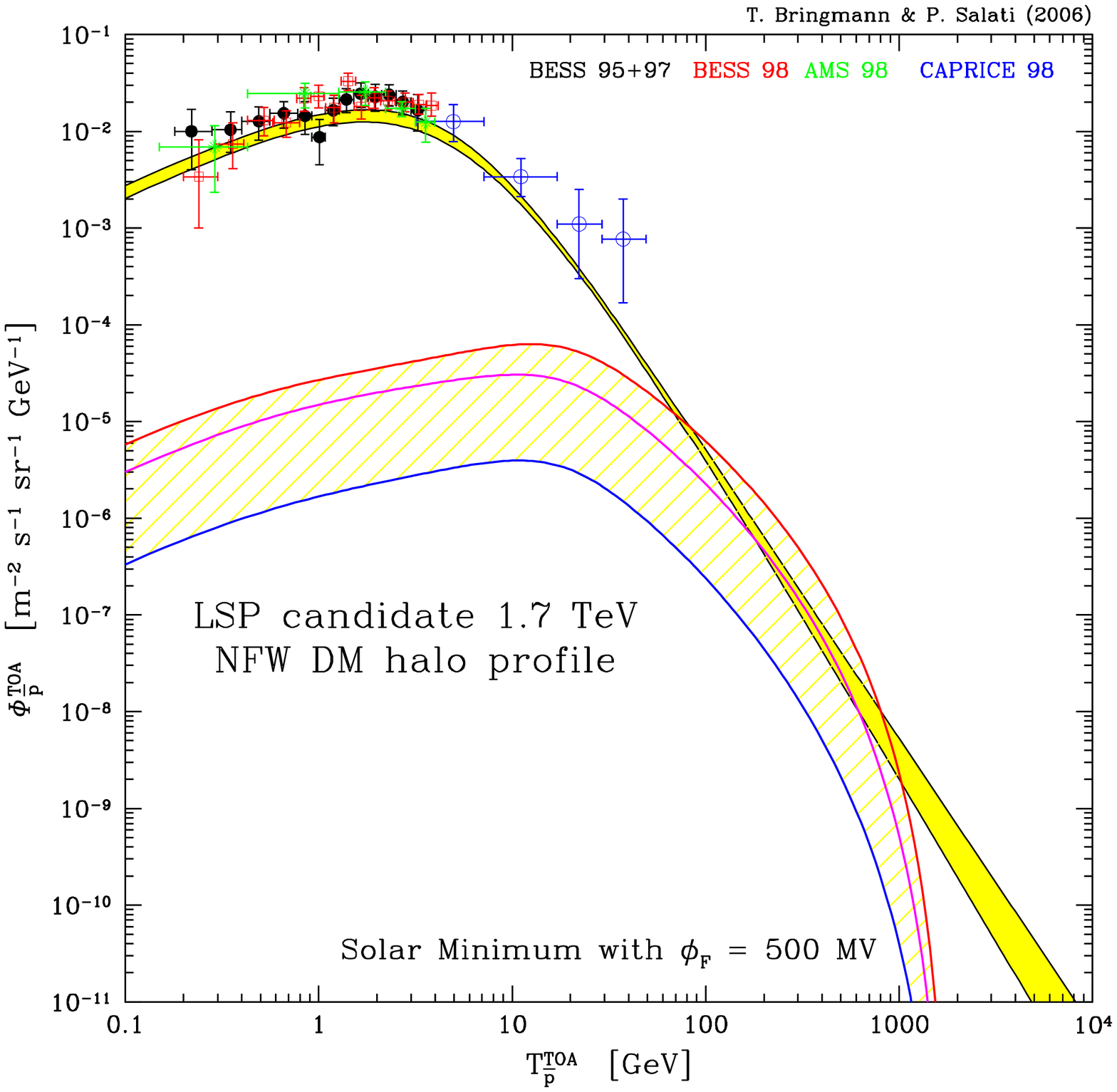}\\
   \end{minipage}  
  \begin{minipage}[t]{0.49\textwidth}
      \centering
  \includegraphics[width=\textwidth]{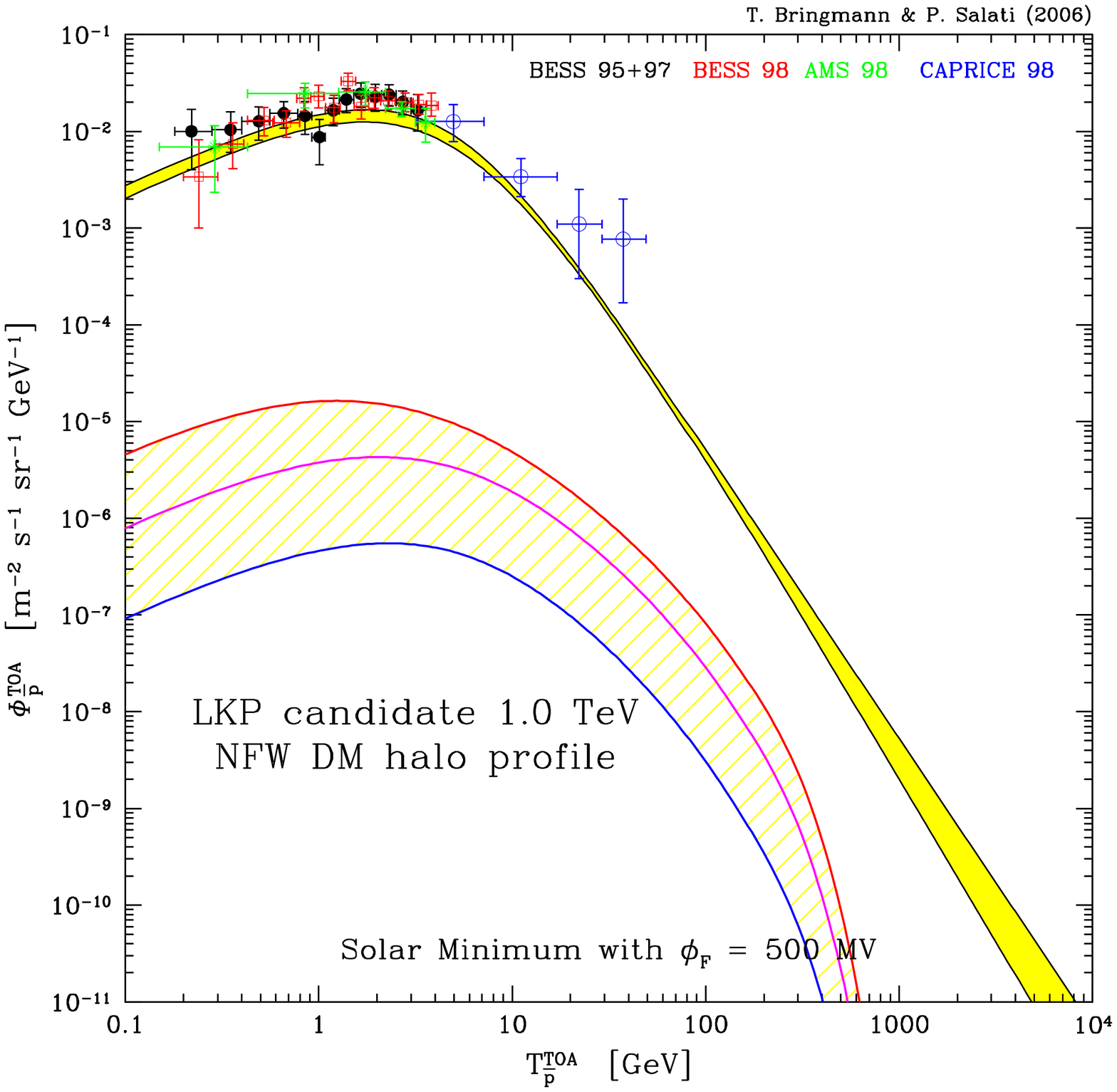}\\
   \end{minipage}
  \begin{minipage}[t]{0.02\textwidth}
   \end{minipage}
  \begin{minipage}[t]{0.49\textwidth}
      \centering   
  \includegraphics[width=\textwidth]{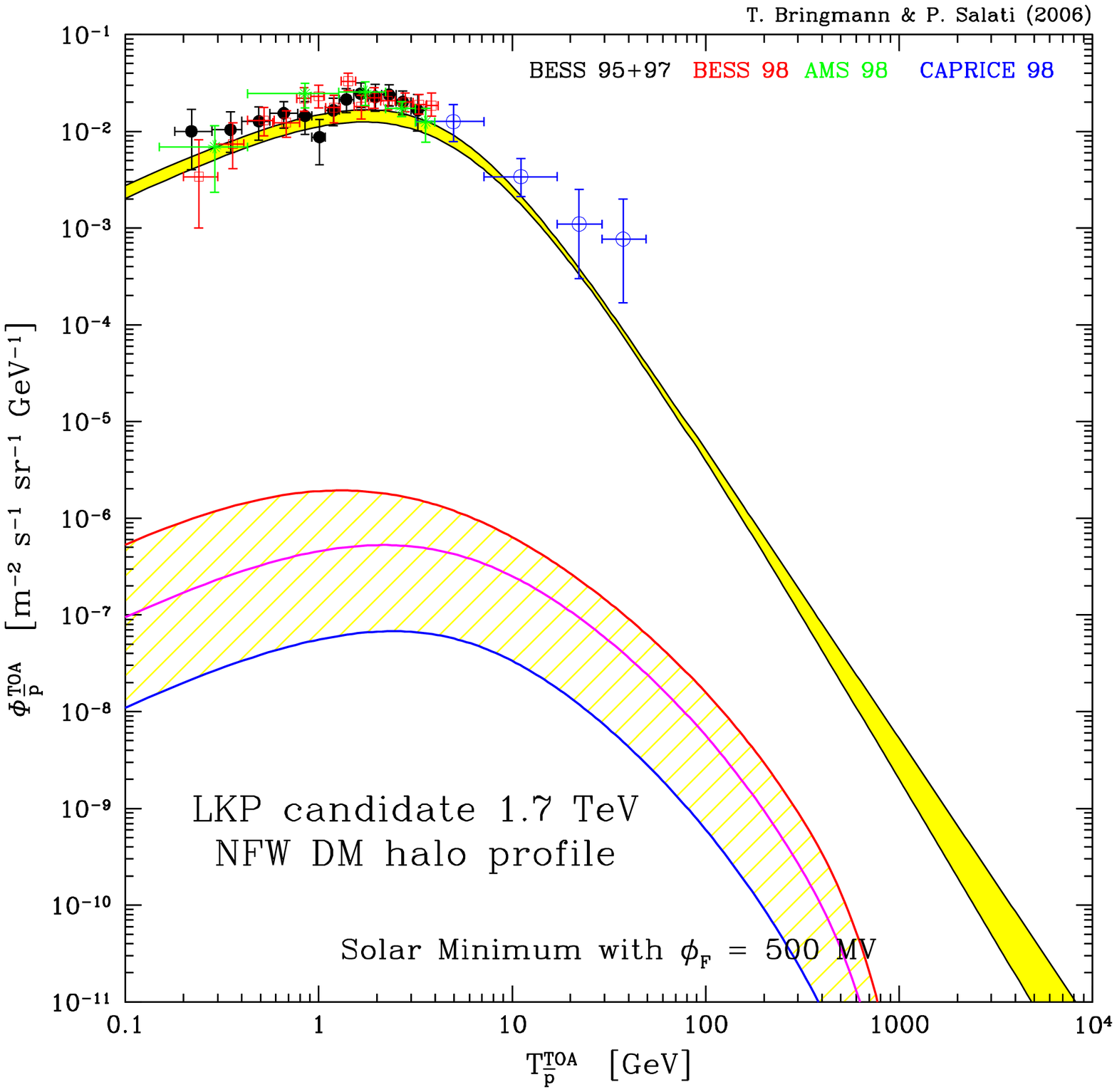}\\
   \end{minipage}
 \caption{Same as Fig. \ref{fig_prof}, now for an NFW halo profile; the diffusion
parameters are varied
%over the whole range that is allowed by the available B/C data.
from the `minimal' to the `maximal' configurations of Table \ref{tab_prop}.}
\label{fig_prop}
\end{figure*}

 With the source function specified in Eq. (\ref{source}), we can easily apply the
formalism described in Section \ref{sec_prop} to compute the corresponding primary flux
of antiprotons. Let us start by showing in Fig.~\ref{fig_prof} and  Fig.~\ref{fig_prop}
the results for our benchmark models, for various halo profiles and diffusion parameters.
Note first that the dependence on the halo profile is as usual much less severe than in
the case of e.g. gamma rays, where the flux usually differs by several orders of
magnitude for the profiles that we consider here; the simple reason for this, of course,
is that the vast majority of antiprotons do not reach far enough through the diffusive
halo to probe the very inner region of the galaxy, where the differences between the halo
profiles are most pronounced. The dependence on the diffusion parameters, on the
contrary, is rather strong -- although the B/C constraints basically fix the secondary
antiproton flux. This effect has been described before \cite{don01,Donato:2003xg} and can be attributed to the fact that
primary and secondary antiprotons mainly probe different regions of the halo. It clearly
illustrates the need for cosmic ray data that are both more accurate and span a larger
energy range, each of which would greatly increase the predictability for primary
contributions to the antiproton flux.

Unfortunately, the expected primary components are usually smaller than the background
fluxes in secondary antiprotons even for favourable diffusion parameters - with the
striking exception of the 1.7 TeV Wino, where the resonantly enhanced annihilation cross
section may allow for a spectacular signal in the range of a few 100 GeV already for very
conservative assumptions about the DM distribution (i.e. no boost factor at all). We can
thus confirm the claim of \cite{his05} that Wino DM exhibits very promising observational
prospects in terms of primary contributions to the antiproton flux.

\begin{figure*}[t]
  \begin{minipage}[t]{0.49\textwidth}
      \centering
  \includegraphics[width=\textwidth]{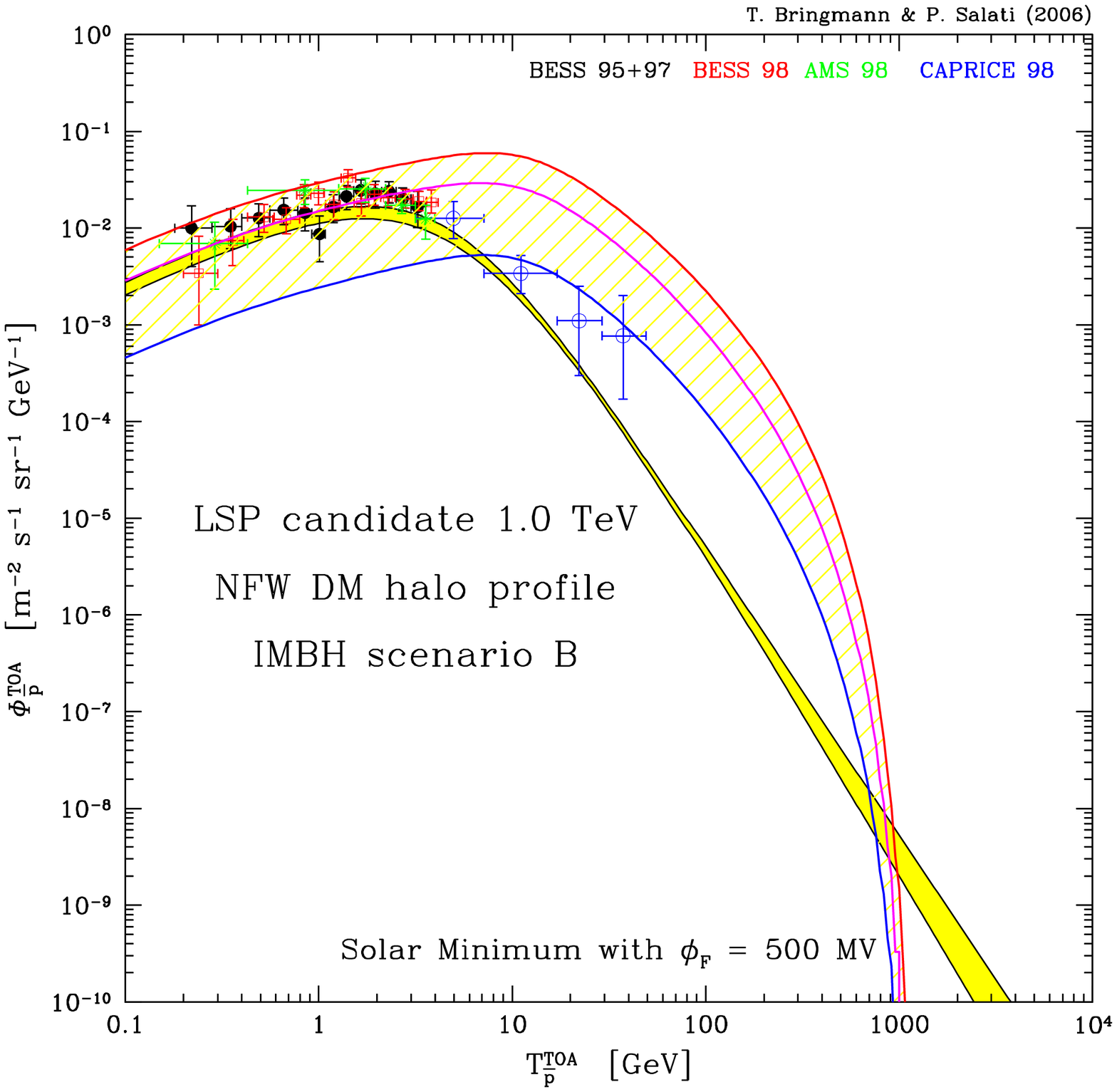}\\
   \end{minipage}
  \begin{minipage}[t]{0.02\textwidth}
   \end{minipage}
  \begin{minipage}[t]{0.49\textwidth}
      \centering   
  \includegraphics[width=\textwidth]{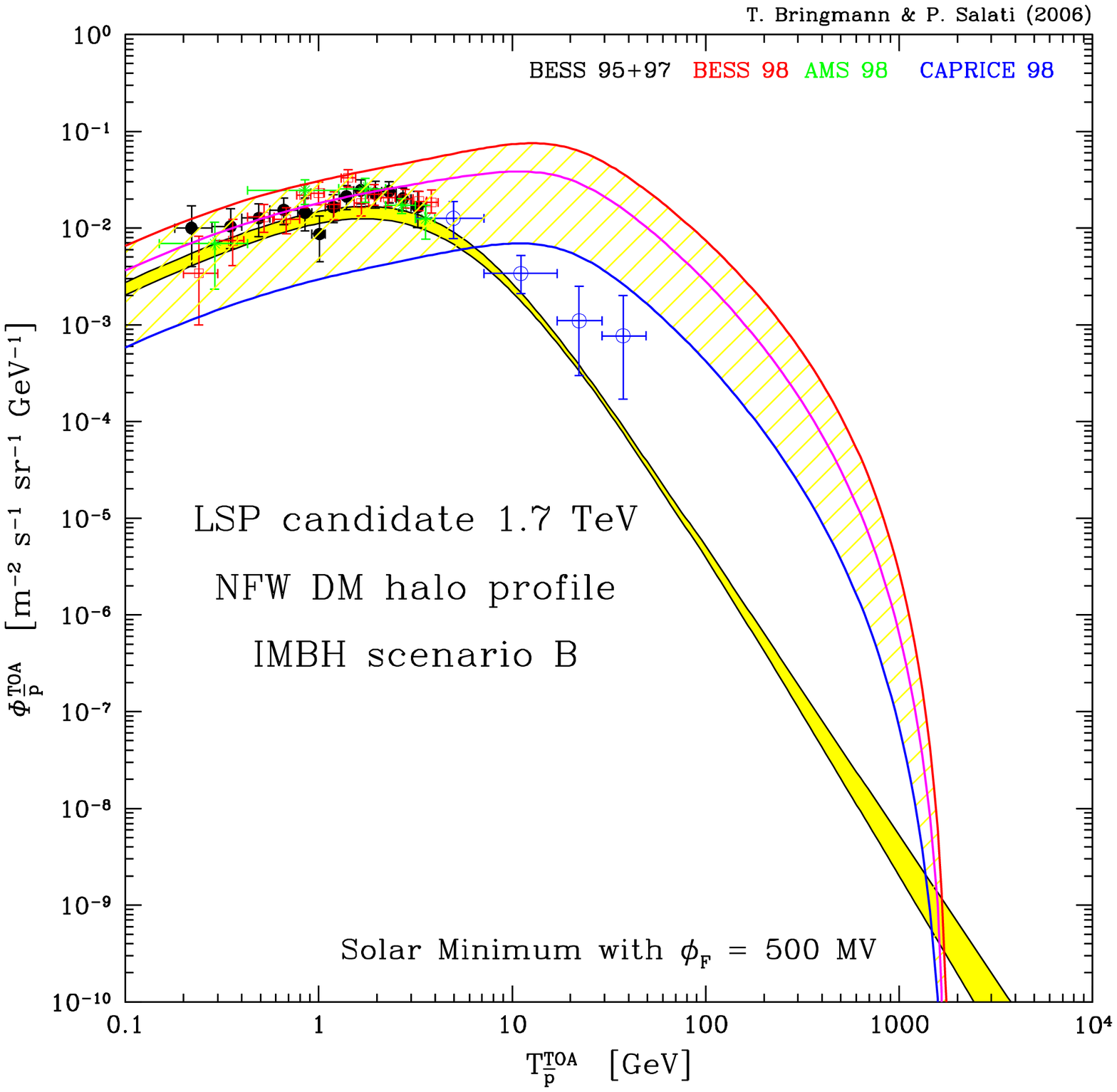}\\
   \end{minipage}  
  \begin{minipage}[t]{0.49\textwidth}
      \centering
  \includegraphics[width=\textwidth]{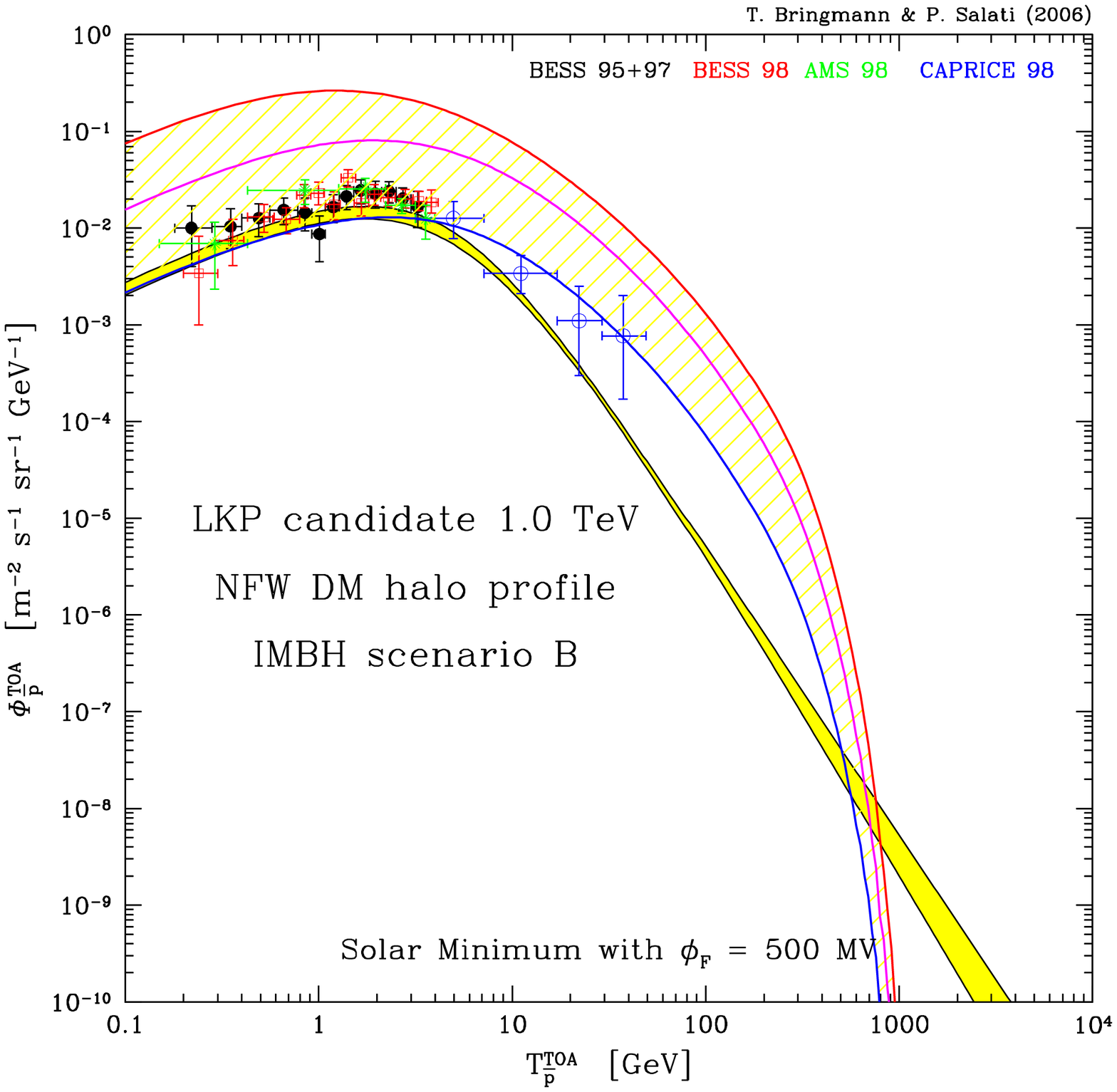}\\
   \end{minipage}
  \begin{minipage}[t]{0.02\textwidth}
   \end{minipage}
  \begin{minipage}[t]{0.49\textwidth}
      \centering   
  \includegraphics[width=\textwidth]{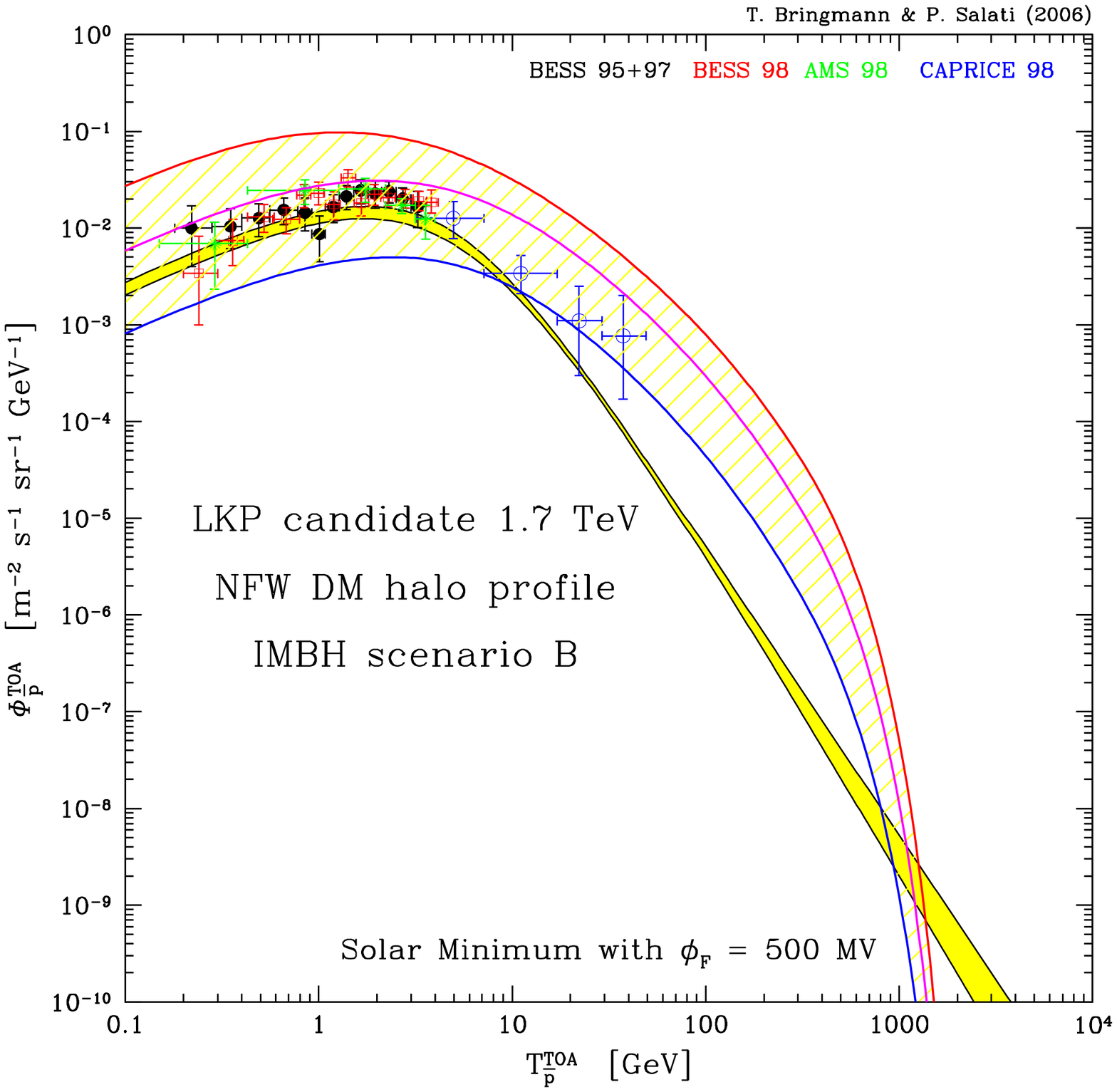}\\
   \end{minipage}
 \caption{Same as Fig. \ref{fig_prop}, now with an NFW profile and a typical population
of IMBHs in the galactic halo.}
 \label{fig_IMBHboost}
\end{figure*}

However, let us now recall from our discussion in the preceeding Section that we actually
do expect primary fluxes that are boosted with respect to what is shown in Figs
\ref{fig_prof} and  \ref{fig_prop}. As we have emphasized, even rather high boost factors
are feasible and appear in several realistic scenarios. To illustrate this point, we
plot in Fig.~\ref{fig_IMBHboost} the primary fluxes for our benchmark models -- as before
for an NFW profile and the diffusion parameter configurations of Table~\ref{tab_prop}, but this time in the
presence of IMBHs. All of a sudden, the situation has drastically changed and a plethora
of new features in the antiproton spectrum seems to lurk just behind the currently
reachable energies~! In fact, with primary fluxes that much enhanced one even has to start
worrying about compatibility with the existing low-energy data; such an analysis, however, would be beyond the scope of the present work and  warrants a
dedicated future study.
\begin{figure*}[t]
  \begin{minipage}[t]{0.49\textwidth}
      \centering
  \includegraphics[width=\textwidth]{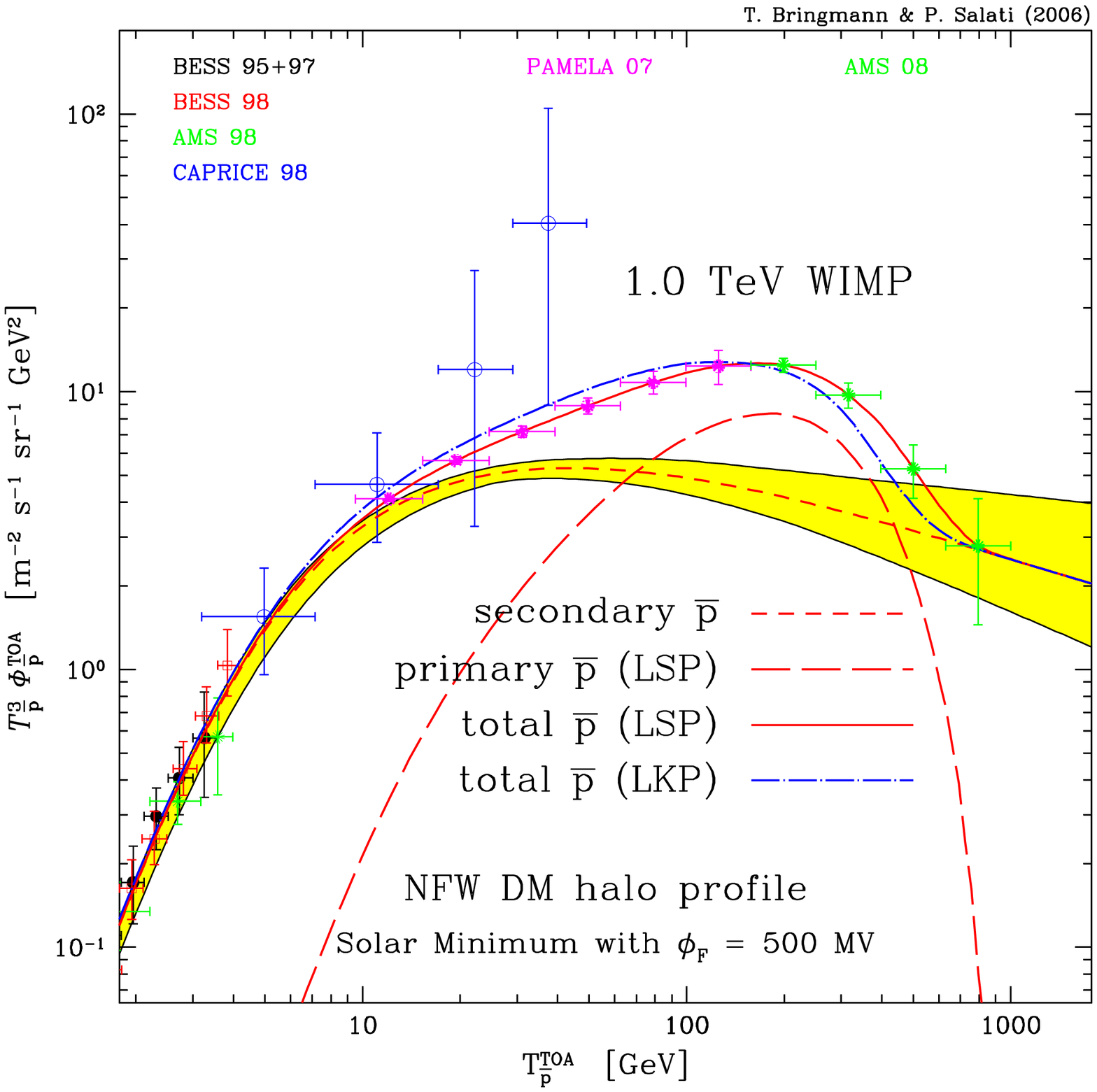}\\
   \end{minipage}
  \begin{minipage}[t]{0.02\textwidth}
   \end{minipage}
  \begin{minipage}[t]{0.49\textwidth}
      \centering   
  \includegraphics[width=\textwidth]{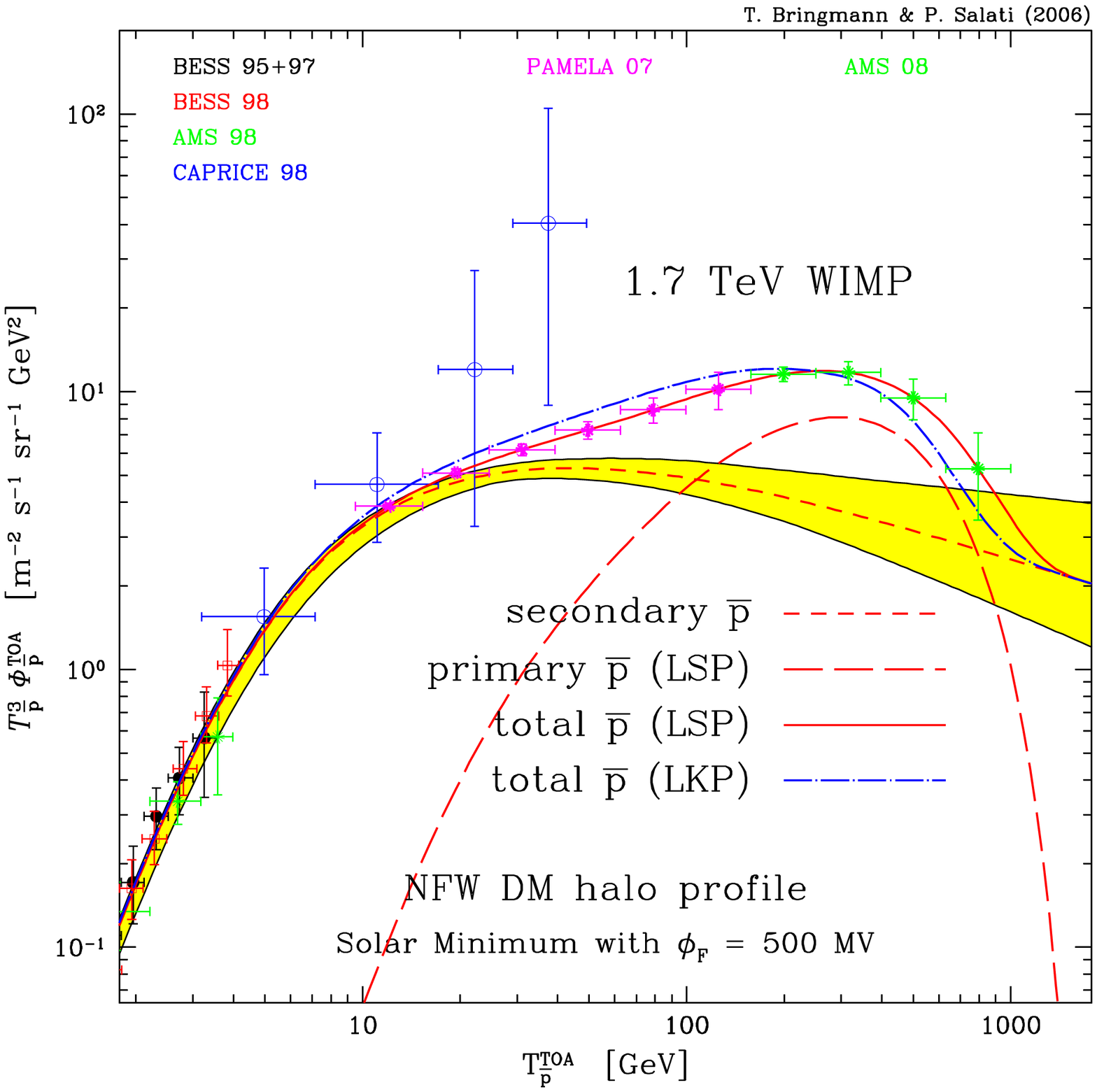}\\
   \end{minipage}  
 \caption{The yellow band shows the expected antiproton background 
for the full range of allowed diffusion parameters.
In the left (right)
panel, we have added the primary and total fluxes for our 1 TeV (1.7 TeV) benchmark
DM models of Table \ref{tab_models}, assuming an NFW halo profile
 and a `medium' set of diffusion parameters; for illustration, and to
 better compare these models, we have adopted a boost factor
of 2 (1050, 330, 270) for the case of an LSP1.7 (LKP1.7, LSP1.0, LKP1.0)
 dark matter
candidate. For the LSP, we also include the expected statistical error
after 3 years of data sampling by PAMELA and AMS-02, respectively. }
 \label{fig_comp}
\end{figure*}

Finally, let us address the question whether the next generation of
 experiments will be able to distinguish between the models that we
 have presented here. To this end, we consider in Fig.~\ref{fig_comp} for
 each model a boost factor that normalizes the maximal deviation from the
 secondary flux to that of the 1.7 TeV Wino case. For the high energies that
 we are interested in, the precision of cosmic ray flux measurements is
 essentially limited by statistics (see also the remark at the end of Section \ref{sec_background}). For comparison, we have therefore included
 in Fig.~\ref{fig_comp} the statistical error after 3 years of data sampling
 by PAMELA and AMS, respectively, provided that these experiments would measure an
 antiproton spectrum as induced by LSP annihilations. As we have seen
 before, a 1.7 TeV LSP can easily be distinguished from the background (even when taking
 into account the full uncertainty in the spectrum of secondary antiprotons) already
 by PAMELA. In fact, this is true  without having to invoke any boost factors at
 all. In order to discriminate the spectra of the other benchmark DM
 candidates at a similar confidence level, one would need boost factor from
 about 150 to 500. While such boost factors may seem rather high from a
 traditional point of view, we have stressed above that, e.g., in scenarios
 with DM mini-spikes around IMBHs one would naturally expect even higher
 values. 

When it comes to the actual discrimination between different DM
 candidates, however, we realize that the prospects are less promising; given
 the current uncertainty in the secondary flux, as well as the expected statistical
 errors in the data, neither PAMELA nor AMS will be able to distinguish between
 different \emph{types} of annihilating WIMPs (i.e.~LSP vs. LKP). A determination
 of the WIMP \emph{mass}, on the other hand, will be possible to a
 certain extent - at least when a clear drop in the spectrum becomes visible
 (for the DM models considered here this could be the case once the AMS data are available). Note also that there appears in the
 spectrum a certain degeneracy between the WIMP type and its mass, putting
 a principle limit on the accuracy of any possible mass determination: for
 a given mass, the LSP produces an annihilation spectrum that is very similar
 to that of an LKP, apart from being slightly
 shifted to higher energies. An LKP with
 a somewhat enhanced mass (by about 10\%) would therefore feature a spectrum that
 is almost indistinguishable from that of the LSP.

%%%%%%%%%%%%%%%%%%%%%%%%%%%%%%%%%%%%%%%%%%%%%%%%%%%%%%%%%%%%%%%%%%%%%%%%%%%%%
\section{Summary
 and Conclusions}
\label{sec_conc}

Upcoming experiments like PAMELA and
 AMS will  extend the upper range of accessible antiproton cosmic ray
 energies considerably, from hitherto around 30 GeV to well above 100
 GeV. In this article, we have considered the galactic antiproton spectrum
 at high energies and performed a detailed analysis of what these -- and
 successive -- experiments are expected to see.
To begin with, we have
 presented the first discussion of the non-trivial behaviour of the
 antiproton \emph{background} at high energies. 
Our
 analysis has been performed in the framework of a two-zone diffusion model, whose parameters can
 be extracted from the spectra of other cosmic ray species (the 
B/C ratio, in particular) in such a way as to give an extremely satisfying explanation
 of the observed low-energy antiproton data {(notice, however, that this is not
 the conclusion reached in \cite{deBoer:2006ck} where galactic diffusion is assumed to be anisotropic).} We have shown that the diffusion
 parameters obtained in this way determine, in fact, almost uniquely the
 spectrum of secondary {antiprotons} up to energies of around 100 GeV. For larger
 energies, the uncertainty in the expected spectrum slowly increases, up
 to a factor of about 6 at 
 10 TeV. The main reason for this uncertainty lies in the fact that the B/C analysis does not sufficiently constrain the spectral index $\delta$ of the diffusion coefficient; this uncertainty will be considerably reduced once experiments like PAMELA and AMS-02 improve the quality and range of existing B/C data. 
Based on both numerical as well as semi-analytical results, we
 furthermore predict a simple power-law scaling for the high-energy secondary spectrum, above slightly less than $100\,$GeV, with
 a spectral index in the range 3.1 to 3.5.

In the second part
 of this paper, we have investigated the possibility of a contribution from
 \emph{primary antiprotons} to the high-energy part of the spectrum,  originating
 from the annihilation of DM particles in the galactic halo. To this end,
 we introduced a set of benchmark models with high DM masses, motivated by
 DM candidates naturally arising in theories with supersymmetric or
 extra-dimensional extensions to the standard model of particle physics.  We
 have shown that the expected primary antiproton signal takes in all
 of these cases a spectral form that is sufficiently different from the
 background to clearly discriminate it against the latter if the corresponding
 total fluxes are high enough. For this latter requirement to be satisfied, one
 usually has to invoke boost factors of $\mathcal{O}(100)$ that at first sight seem rather high
 (with the interesting exception of a Wino-like DM candidate, see also \cite{his05}). As a
 side-result of the present work, however, we have also derived the expected boost factors
 in the IMBH scenario of \cite{imbh} and demonstrated that in this case, 
in fact, they are even higher than what is needed for a clear attribution of
 the observed signal to a DM (as opposed to secondary antiproton) origin.
We, furthermore, note that the most recent, high-resolution simulations of gravitational clustering \cite{Diemand:2006ik} hint at a considerable DM fraction to be distributed in clumps of highly enhanced DM densities, which potentially would lead to large boost factors, too.

 When it comes to the possibility of distinguishing between different DM
 candidates, however, the antiproton spectrum does not turn out to be very
 well suited and other indirect DM detection methods (gamma-rays, in
 particular) are probably more promising in that respect. The mass of
 the annihilating DM particle, on the other hand, may eventually be
 determined, to a fair accuracy, from the antiproton spectrum alone; a comparison with
 the result of other DM searches would then give an important,  
independent piece of information. It is also worthwhile to study in this context the
 cross-correlation of antiproton signals with DM-induced features in
 the spectrum of other charged cosmic ray species (for the LKP, e.g.,
 this would be a pronounced peak in the \emph{positron} spectrum that necessarily has to appear if the antiproton spectrum shows an LKP-induced distortion \cite{Hooper:2004xn,bri05} -- a feature that is not shared by LSP DM candidates).

To conclude, measurements
 of the galactic antiproton spectrum are an important testbed for our
 understanding of the propagation of charged particles through the galaxy
 and the diffusive structure of the Milky Way halo. As a means of indirect
 DM detection, antiprotons are probably not the single-most promising species;
 we would like to stress, however, that one should, rather, focus on
 the complementarity of different approaches, to which antiproton measurements
 would then have the potential of providing an important contribution.

%%%%%%%%%%%%%%%%%%%%%%%%%%%%%%%%%%%%%%%%%%%%%%%%%%%%%%%%%%%%%%%%%%%%%%%%%%%%%
\begin{acknowledgments}
\mbox{ }\vspace{-4ex}\\

It is a pleasure to thank Pierre Brun for many interesting discussions on antiproton fragmentation functions in the early stages of this work.
We are, furthermore, grateful to Laurent Der\^{o}me who helped us with the
antiproton differential production cross section in $p + A$
interactions and to Joakim Edsj\"o for support with the \textsc{DarkSUSY} package. 
P.S. would like to thank the french programme national de cosmologie
PNC for its financial support. T.B. warmly acknowledges the hospitality and pleasant atmosphere of 
the Laboratoire d'Annecy-le-Vieux de Physique Th\'eorique (LAPTH), where part of this work was performed.
\end{acknowledgments} 

%%%%%%%%%%%%%%%%%%%%%%%%%%%%%%%%%%%%%%%%%%%%%%%%%%%%%%%%%%%%%%%%%%%%%%%%%%%%%

%%%%%%%%%%%%%%%%%%%%%%%%%%%%%%%%%%%%%%%%%%%%%%%%%%%%%%%%%%%%%%%%%%%%%%
%%%%%%%%%%%%%%%%%%%%%%%%%%%%%%%%%%%%%%%%%%%%%%%%%%%%%%%%%%%%%%%%%%%%%%
\end{document}